\documentclass[aps,pre,showpacs,groupedaddress]{revtex4}
\usepackage{amsfonts}
\usepackage{mathrsfs}
\usepackage{graphicx}
\usepackage{amsmath}
\usepackage{amssymb}
\usepackage{amsbsy}
\usepackage{marvosym}
\usepackage{url}
\makeatletter
\newcommand{\linea}{\noindent\rule{1.0\textwidth}{1pt}}%

\newenvironment{algo}{\medskip\noindent\begin{minipage}{1.0\textwidth}\linea\begin{sf}\\}{\end{sf}\linea\end{minipage}\medskip}

\usepackage{color}
\makeatother

\begin{document}

\title{The network source location problem: ground state energy, entropy and effects of freezing}

\author{Haiping Huang$^{1}$, Jack Raymond$^{2}$ and K. Y. Michael Wong$^1$}
\affiliation{$^1$Department of Physics, The Hong Kong University of
Science and Technology, Clear Water Bay, Hong Kong, China\\
$^2$Dipartimento di Fisica, Universit\`a La Sapienza, Piazzale Aldo
Moro 5, I-00185 Roma, Italy}
\date{\today}

\begin{abstract}
Ground state entropy of the network source location problem is
evaluated at both the replica symmetric level and one-step replica
symmetry breaking level using the entropic cavity method. The regime
that is a focus of this study, is closely related to the vertex
cover problem with randomly quenched covered nodes. The resulting
entropic message passing inspired decimation and reinforcement
algorithms are used to identify the optimal location of sources in
single instances of transportation networks. The conventional belief
propagation without taking the entropic effect into account is also
compared. We find that in the glassy phase the entropic message
passing inspired decimation yields a lower ground state energy
compared to the belief propagation without taking the entropic
effect. Using the extremal optimization algorithm, we study the
ground state energy and the fraction of frozen hubs, and extend the
algorithm to collect statistics of the entropy. The theoretical
results are compared with the extremal optimization results.
\end{abstract}

\pacs{64.60.aq, 02.50.Tt, 64.70.P-, 89.20.-a}
\keywords{cavity method, disordered systems (theory), entropy,
message-passing algorithms, transportation networks}
 \maketitle

\section{Introduction}
Statistical physics methods and ideas inherited from studies of
disordered systems play an important role in providing theoretical
insights and developing low complexity algorithms for combinatorial
optimization problems or constraint satisfaction
problems~\cite{cavity-2001,cavity-2003,MM-2009}. These problems
become subjects of interest across a variety of different
disciplines such as computer science, discrete mathematics,
statistical physics, engineering and computational
biology~\cite{MM-2009}. One archetype that involves both continuous
and discrete variables is the network source location problem where
we try to find optimal location of sources in a transportation
network with an optimal flow pattern. Studies on this source
location problem~\cite{Wong-2006,Wong-09pre,Wong-09jstat,Wong-2010}
are practically relevant to network design and
optimization~\cite{CH-12} and can have widespread applications in
the field of operations research~\cite{Rardin-98}.

In the source location problem, transportation networks are assumed
to be composed of nodes with either a surplus or a deficiency of
resources. How to distribute the resources and to replace some
deficient nodes by resource providing ones becomes an important
problem in network optimization. After the optimization, the
remaining deficient nodes serve as consumer nodes and a network-wide
satisfaction is achieved. A series of discontinuous transitions with
different configurations of the source and consumer
nodes~\cite{Wong-09pre,Wong-2010} can be observed by varying model
parameters. Two of them are the singlet regime where the consumer
nodes are isolated and the doublet regime where the consumer nodes
can be paired or isolated. Even in the singlet regime, there exist a
glassy phase in which optimizing the location of sources becomes
algorithmically hard, which is most likely due to the emergence of
long range frustrations~\cite{ZH-prl05,ZH-njp05}.

In this work, we apply the entropic cavity method to evaluate the
typical value of the ground state entropy in the singlet regime and
probe the entropic effects. The entropic cavity method has been used
to compute the ground state entropy of the minimal vertex cover
(MVC) problem~\cite{ZH-pre09} and to study entropic effects in the
channel coding problem and budget-constrained
auctions~\cite{Huang-pre09,Zecchina-09}. The purpose of this paper
is to consider the benefits brought by including the entropic
effects properly, especially in terms of solutions (optimal
assignments of source locations) found by improved message passing
algorithms. Hereafter, we identify an assignment to be optimal if it
yields the minimal energy cost in a specific regime (e.g., the
singlet or doublet connection pattern). For verification, a standard
implementation of the extremal optimization~\cite{Boet-2001} is used
to study the energetic properties, and this is adapted as a biased
sampling method for the ground states which appears for small
systems sufficient to enumerate all the ground states and frozen
variables.

 The rest of this paper is organized as follows. The source location problem is defined
in detail in Sec.~\ref{sec_SLP}. Following this definition we
provide more details on the related literatures and previous
results. The analysis at both the replica symmetric (RS) level and
one-step replica symmetry broken ($1$RSB) level by the entropic
cavity method is presented in Sec.~\ref{sec_ECM}. The extremal
optimization (EO) is developed; numerical and theoretical results
are compared. In Sec.~\ref{sec_result}, experimental studies of the
proposed maximal decimation and reinforcement strategy are carried
out. Conclusion and some future directions are given in
Sec.~\ref{sec_Sum}. Some technical details are given in the
Appendices.

\section{The source location problem}
\label{sec_SLP}

The facility location problem is an important problem in management
science, since the placement of facilities at optimal positions in a
network is able to provide efficient services while minimizing the
logistics cost~\cite{Revell-1977}. The services may be public ones
that concern the well-being of all citizens, such as ambulance
service or public schools, or they may be private ones that concern
the sphere of influence and maintenance cost of retailing
companies~\cite{Revell-1986}. Similar issues exist in wireless
sensor networks~\cite{Karaki-2004,Frey-2009}, in which sensors are
deployed in an arena for purposes such as surveillance, fire
detection, and collection of meteorological and pollution data. Due
to the limited battery size of the sensors, minimizing the
transportation cost becomes an essential issue to prolong the life
span of the networks. However, the traditional approach to the
problem is integer programming, whose complexity scales up rapidly
with system size, and solutions for large-scale systems depend on
heuristics~\cite{Pirkula-1998}.

In this paper, we consider a transportation network of $N$ nodes in
which resources are transported through the links, so that the
resource demands of all nodes are satisfied. There are two kinds of
nodes. The {\em surplus nodes} supply the resources and the {\em
deficient nodes} have demands for resources. To the {\em deficient
nodes} $i$ is associated a capacity $\Lambda_i=-1$, corresponding to
the amount of resources consumed by $i$, and to the {\em surplus
nodes} $i$ is associated a sufficiently large capacity $\Lambda_i=
A\gg1$, corresponding to the amount of resources it can provide.
This scenario is typical in many modern applications of network
traffic optimization, such as logistic networks and sensor networks,
where the surplus nodes represent distribution centers in logistic
networks or base stations in sensor networks.

The version of the source location problem considered in this paper
is an optimization problem in the space of real-valued variables
$x_{ij}$ and the Boolean variables $s_{i}$. The variables $s_i$
defined over the set of deficient nodes $\mathcal{D}$ are indicator
functions ($1=$ true, $0$ false) for an installation, which is a
reassignment of a deficient node capacity from $-1$ to $+A$ (making
it behave as a surplus node). For each link we also define a
real-valued flow $x_{ij}$ ( $=-x_{ji}$) from node $j$ to $i$. Those
deficient nodes not reassigned ($s_i=0$) will be called {\em
consumer nodes} and will have a net inward flow; all other nodes are
called {\em source nodes} and will have zero, or positive net
outward flow. In a valid assignment either a deficient node $i$ must
be installed as a source node ($s_i=1$), or the flows are required
to satisfy the non-negativity constraints of the final resource,
defined by $\xi_{i}$ for every node, i.e.,
\begin{equation}
    \xi_{i}\equiv \Lambda_{i}  + \sum_{j\in\partial i}x_{ij}\geq0,
\label{fres}
\end{equation}
where  $\partial i$ denotes neighbors of node $i$. To valid flows
and installations is further associated the energy
\begin{equation}
    E=\sum_{\left(ij\right)}\frac{x_{ij}^{2}}{2}+\frac{u^{2}}{2}\sum_{i\in\mathcal{D}}s_{i}.
    \label{cost}
\end{equation}
The task will be to optimize over valid flows and installations to
minimize this quantity. The first term is the total transportation
cost summed over each link $(ij)$.  The transportation cost on link
$(ij)$ is quadratic in the flow. The quadratic function is chosen
because of its convex property, which tends to balance the traffic
load among the links; other convex functions can work equally
well~\cite{Banavar-2000,Shao-2007,Bohn-2007}. The second term is the
installation cost of converting a deficient node to a source node.

We will be interested in studying an ensemble of diluted networks in
which each node has the same degree $C$. The locations of the
surplus and deficient nodes are random, with a fraction of surplus
nodes given by $\phi_s$, hence capacities are assigned uniformly at
random according to the distribution
$\rho(\Lambda)=\phi_{s}\delta(\Lambda-A)+(1-\phi_{s})\delta(\Lambda+1)$.
Other ensembles are certainly interesting, and our analysis can be
easily extended to networks with fluctuating degrees (e.g.,
Erd\"os-R\'enyi random networks), or with other distributions of
$\Lambda$. Many interesting phenomena can be exactly analyzed in
this restricted setting, with implications for technologically
relevant large networks.

\subsection{Known results and developments with respect to related models}
It has been previously shown that the optimal set of suppliers
(source locations) can be obtained by assuming that all deficient
nodes are consumer nodes, and then optimizing the nonlinear energy
cost function in the space of the real-valued variables $x_{ij}$
given by~\cite{Wong-09pre,Wong-2010}
\begin{equation}
    E = \sum_{\left(ij\right)}\frac{x_{ij}^{2}}{2}+\frac{u^{2}}{2}\sum_{i\in\mathcal
    {D}}\Theta\left(-\xi_{i}\right),
\label{cost2}
\end{equation}
 where $\Theta(x)=1$ if $x>0$ and
$\Theta(x)=0$ otherwise. Then we identify those nodes with
non-negative final resources as the consumer nodes, and naturally
assign those nodes with negative final resources to be the source
nodes in the optimal solution of the source location problem.

\begin{figure}
         \includegraphics[bb=99 75 502 224,scale=0.8]{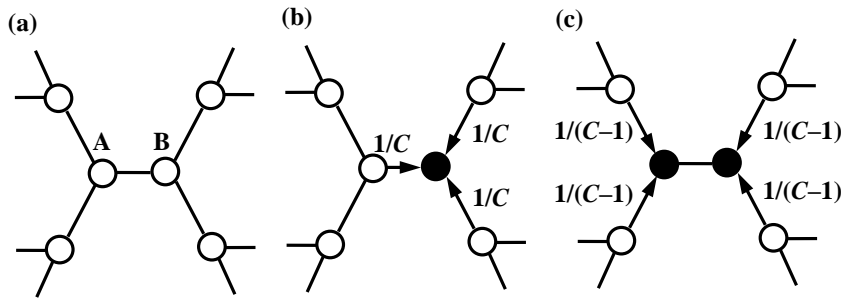}
  \caption{ Illustrations showing that two nodes labeled $A$ and $B$ in a transportation network are energetically more stable in the singlet state
  than in the all-source state or the doublet state. White and black nodes are source and consumer states respectively.
  Arrows indicate flows of magnitude $1/C$ or $1/(C-1)$. (a) The all-source state in which the energy of nodes $A$ and $B$ is $2(u^{2}/2) = u^{2}$. (
  b) The singlet state with energy $u^{2}/2 + (C/2)(1/C)^{2}$. (c) The doublet state with energy $2[(C-1)/2][1/(C-1)]^{2}=1/(C-1)$.
   }\label{ebpfig1}
 \end{figure}

The ground state energy of this model can be analyzed by the cavity
method at zero temperature. Previous
works~\cite{Wong-2006,Wong-2007,Wong-09jstat,Wong-2010} showed that
the cavity energy functions under Eq.~(\ref{cost2}) with continuous
variables can be decomposed into composite functions parameterized
by their energy minima, such that the recursions of cavity energy
functions can be converted to simple recursions of probabilities,
which simplifies the analysis a lot. This simplification is due to
the quadratic form of the transportation cost, and thus the
composite function captures the multi-valley features of the cavity
energy functions. The authors in Refs.~\cite{Wong-09pre,Wong-2010}
further found that as the installation cost parameter $u$ changes,
different configurations of consumer and source nodes appear. This
leads to a cascade of phase transitions in the glassy phase with
abrupt jumps of the fraction of source nodes in the optimized
network. For instance, in the singlet regime where
$1/\sqrt{C}<u<\sqrt{(C+1)/[C(C-1)]}$, the consumer nodes are
isolated. This is because in this range, the singly consuming state
is always energetically more stable compared with its two
neighboring phases, i.e.,  all-source phase and doublet phase, as
derived in Refs.~\cite{Wong-09pre,Wong-2010} and illustrated in
Fig.~\ref{ebpfig1} ($C=3$). The simple flow configuration in the
singlet regime (see also Appendix~\ref{sec:CAV}) allows us to
determine the optimal solution easily. The flow in a link into a
consumer node is always $1/C$, and the flows in other links are $0$.
Hence the flow configuration is determined once the nodes are
determined to be in the source state or the consumer state. This
further simplifies the source location problem into a
discrete-valued optimization.

\begin{figure}
         \includegraphics[bb=145 76 466 324, scale=0.8]{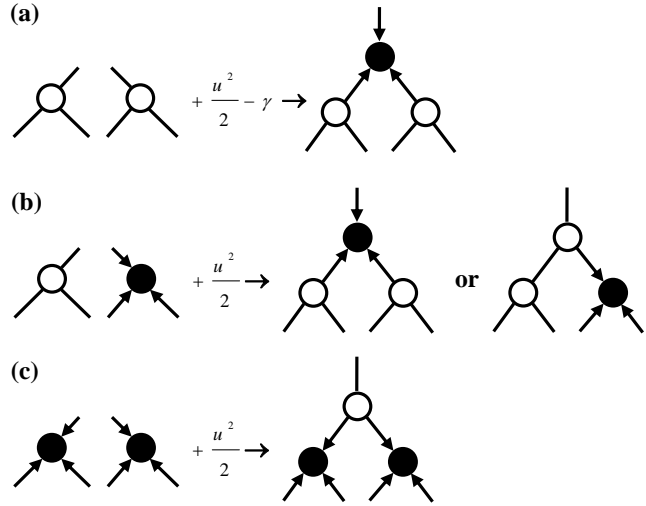}
  \caption{ Recursion relations of the cavity states of deficient nodes.
  White and black nodes represent cavity source and cavity consumer states respectively.
  Arrows represent flows of magnitude $1/C$ when (a) there are $C-1$ cavity source states,
  (b) one of the $C-1$ cavity states is a consumer, (c) two or more of the $C-1$ cavity states are consumers.
  The cavity states of surplus nodes are always the source state.
   }\label{ebpfig2}
 \end{figure}

Based on the cavity method, the recursion relations of the cavity
source and cavity consumer states (more details will be given in
Sec.~\ref{sec_ECM}) are derived in Appendix~\ref{sec:CAV}. The
cavity energy of the {\em c}-state relative to the {\em s}-state is
given by
\begin{equation}\label{recSimp}
    \epsilon_{i\rightarrow j}=-\gamma-\sum_{k\in\partial i\backslash j}\min(0,\epsilon_{k\rightarrow i}),
\end{equation}
where $\gamma\equiv u^{2}/2-1/(2C)>0$ in the singlet regime. The
simple recursion leads to $\epsilon_{i\rightarrow
j}=k\gamma(k=-1,0,\ldots,C-2)$ with $k=-1$ for the cavity consuming
state and $k=0$ for the cavity bistable state and others for the
cavity resource providing state. This equation tells us that a node
$i$ is in the {\em c} state if its neighbors except node $j$ are all
in the {\em s} state, and $-\gamma$ gives the relative cavity energy
to the {\em s} state, which is exactly what we shall use in
Sec.~\ref{RS01} and Sec.~\ref{RS03}. The full state of a node $i$,
after taking into account the cavity states of all neighbors, is
given by
\begin{equation}\label{recSimpfull}
    \epsilon_{i}=-\gamma-\sum_{k\in\partial i}\min(0,\epsilon_{k\rightarrow
    i}).
\end{equation}
The results are summarized in Fig.~\ref{ebpfig2} ($C=3$). While the
results are apparently intuitive from Fig.~\ref{ebpfig2}, we remark
that they are rigorously based on the cavity derivation described in
Appendix~\ref{sec:CAV}.

These results give rise to a belief propagation
algorithm~\cite{Wong-09pre,Wong-2010}. When the fraction of surplus
nodes is sufficiently large, it converges satisfactorily and
provides excellent agreement with the simulation results in terms of
the fraction of source nodes. However, the algorithm was less
satisfactory when the fraction of surplus nodes is not large. This
is the regime where the cavity recursive relations become unstable
towards fluctuations or, in the framework of the replica method,
where the replica symmetric solution becomes unstable.

 One drawback of the previous analysis is that the degeneracy of the
cavity source state and the cavity consumer state has been ignored.
Algorithmic hardness has also been studied, and is associated with
glassy behavior in thermodynamics, which arises from the assignment
of nodes with freedom to be resource providers or consumers. These
nodes are called bistable nodes (Fig.~\ref{ebpfig2} (b), see also
Ref.~\cite{Wong-2010}), and often exist in chains, so that when one
node in the singlet regime is assigned to be a consumer, its
bistable neighbor is then required to be resource providing, and its
next nearest neighbor a consumer, and so on. The correlations of
state assignments may be rather long ranged. Without the information
about the entropy of the respective states, random assignments of
the bistable nodes typically causes contradictions throughout the
network. That is to say, for the bistable nodes, the source and
consumer states may not occur with equal probability due to entropic
effects, even though their energy contribution is the same. In
Ref.~\cite{Wong-2010}, the cavity bistable state and the cavity
source state were grouped together in deriving their recursion
relations. Thus the entropy of the ground state was not considered
properly. In this paper, we will compute the entropy of the ground
state extending the previous efforts~\cite{Wong-2010,ZH-pre09},
restricting our attention to the {\em singlet} regime. We will also
propose the entropic message passing inspired decimation algorithm
and reinforcement strategy to identify the optimal location of
sources in both the easy (replica symmetric) and hard (replica
symmetry breaking) phases.

The singlet regime is the focus for application of the methods we
develop. In this regime, the source location problem has an
interesting connection with the MVC problem~\cite{ZH-pre06,ZH-pre09}
if we assume the source nodes as the covered nodes, and the consumer
nodes as the uncovered nodes. The difference is the existence of the
surplus (quenched to the covered state) nodes as the quenched
disorder in the source location problem.

We can see this correspondence by assigning energies $0$ and $1$
respectively to the covered and uncovered states of a node. Then the
cavity recursion relation of the MVC problem can be written as
\begin{equation}\label{mvc}
    \epsilon^{{\rm MVC}}_{i\rightarrow j}=\max\left[1-\sum_{k\in\partial i\backslash j}\epsilon_{k\rightarrow i}^{{\rm
    MVC}},0\right].
\end{equation}
Introducing the transformation $\epsilon^{{\rm MVC}}_{i\rightarrow
j}=\max(-\epsilon_{i\rightarrow j}/\gamma,0)$, we can verify that
the recursion relation of the source location problem implies
Eq.~(\ref{mvc}). On the other hand, the full state of node $i$ in
the MVC problem, after taking into account the cavity states of all
neighbors, is given by
\begin{equation}\label{mvcfull}
    \epsilon^{{\rm MVC}}_{i}=\max\left[1-\sum_{k\in\partial i}\epsilon_{k\rightarrow i}^{{\rm
    MVC}},0\right].
\end{equation}
Note that $\epsilon^{{\rm MVC}}_{i}$ only takes the values $0$ and
$1$. In contrast, $\epsilon_i$ in Eq.~(\ref{recSimpfull}) takes the
values $k\gamma(k=-1,0,\ldots,C-1)$. This implies that the solutions
of the source location problem and the MVC problem are identical
only in the ground state. Their excited states yield different
energies. As we shall see, this will lead to different estimates of
average energies in the $1$RSB ansatz, since the reweighting factors
in the $1$RSB picture depend on the energies of the excited states.

Despite the similarities of the two problems, this paper is
motivated by further considerations in the following aspects. First,
our inclusion of surplus nodes as quenched variables arises
naturally from realistic considerations. In real applications,
networks often consist of already installed facilities and the
objective is to install further facilities for service improvement;
there is no point to demolish the existing facilities. Second, we
will show that care has to be taken in deriving the recursion
relations of the cavity probabilities and the expression of the free
energy per node. This is because the energy in the source location
problem is distributed both among the nodes and links, while the
energy of the minimal vertex cover problem is only defined in terms
of the states of the nodes. As we shall see, this leads to two kinds
of cavity analysis, to be referred to as {\em reconnection} and {\em
restoration} (Sec.~\ref{RS03} and Appendix~\ref{app:RS02}). No such
subtleties exist in the vertex cover problem.

Compared with previous work calculating the ground state entropy of
the vertex cover problem~\cite{ZH-pre09}, we will show that the
energetic consideration is necessary when we consider replica
symmetry-breaking effects, for which metastable states have to be
weighted by energetic reweighting factors depending on flows on the
links. Furthermore, for future generalizations to more complex
scenarios (such as the doublet regime), it is more convenient to
develop the method with consideration of the energetic influence of
links (see also a brief description in Appendix~\ref{sec:CAV}).
Phenomenologically, we will show that the inclusion of quenched
surplus nodes produces a rich picture of the optimal states,
illustrated by the presence of frozen nodes, and the roles played by
frozen hubs and peripheral nodes observed in extremal optimization,
and moreover an improved message passing algorithms applicable to
single instances.

\section{The entropic cavity method}
\label{sec_ECM}

In this section, we will present the entropic cavity method to
compute the ground state entropy of the source location problem at
the replica symmetric level and one-step replica symmetry breaking
level. At the replica symmetric level, there exists a single ground
state and the clustering hypothesis that the correlation between any
two randomly selected nodes in a large sparse network is weak
becomes valid. We derive a closed set of equations involving two
messages (cavity probability and cavity entropy). The replica
symmetric ground state entropy can be evaluated from the fixed point
of the recursive equations. Extending the analysis to one-step
replica symmetry broken level is straightforward. When replica
symmetry is broken, the single ground state would split into
exponentially many states, which violates the clustering hypothesis.
Although the weak correlation assumption is still satisfied in each
state, the energy level crossings of states under the cavity
iterations should be taken into account. All mean-field analysis
presented here are restricted to the zero temperature limit which
selects the ground state.

\subsection{Replica symmetric analysis}
\label{sec:RS}
\subsubsection{Recursion relation}
\label{RS01}
 Under the replica symmetric approximation, the joint
state distribution of any two randomly chosen nodes ($C=3$) from the
large diluted network takes a factorized form making the derivation
of a recursive relation feasible. Applying the cavity method to the
source location problem, we consider the state of a node $i$ in the
absence of one of its neighbors $j$. In the singlet regime, it can
be either in the cavity consumer state, or the cavity source state,
denoted by {\em c} and {\em s} respectively. In the {\em c} state,
the cavity energy of a node is lowest when a flow of $1/C$ enters
it, whereas in the {\em s} state, the cavity energy of the node is
lowest when no flow enters it (see also Fig.~\ref{ebpfig2}).

Let us first define $\psi_{i\rightarrow j}^{s}$ as the cavity
probability that node $i$ is in the {\em s} state in the absence of
node $j$ when the flow on the forward link $i\rightarrow j$ is
included in calculating the optimal state. It can also be viewed as
the message passing from the node $i$ to node $j$. If
$\psi_{i\rightarrow j}^{s}=1$, we say that node $i$ takes the {\em
s} state in the absence of node $j$. $\psi_{i\rightarrow j}^{s}=0$
indicates node $i$ should take the {\em c} state without node $j$.
Otherwise, if $\psi_{i\rightarrow j}^{s}\in (0,1)$, then the {\em s}
and {\em c} states of node $i$ are degenerate without node $j$.

To derive a formula for the ground state entropy, a pair of messages
(cavity probability and cavity entropy) will be involved. Below, we
derive the recursion of $\psi_{i\rightarrow j}^{s}$ and the entropy
change by considering the energetics of both the nodes and links for
the source location problem in the zero temperature limit.
Alternatively, the recursion relation can be derived by focusing on
the cavity states of the nodes only. This is described in the
entropic derivation in Appendix~\ref{sec:appendix} and the replica
symmetric entropy formula is the same as that for the vertex cover
problem~\cite{ZH-pre09,Pan-2009}. However, the entropic derivation
cannot provide the energy changes in the recursive steps, which are
required to calculate the reweighting factors in the $1$RSB
analysis. In the source location problem, we should take the forward
link into account to derive the cavity energy and entropy values.

A node $i$ is in the {\em c} state if its neighbors except node $j$
(denoted by $\partial i\backslash j$) are all in the {\em s} state,
and its cavity energy relative to the {\em s} state is $-\gamma$. On
the other hand, node $i$ can take the {\em s} state for any
combination of states of the neighbors in the set $\partial
i\backslash j$ (see Fig.~\ref{ebpfig2}). Hence, we can write the
following cavity free energy
\begin{equation}\label{cavfe01}
        F_{i\rightarrow j}=\frac{u^{2}}{2}-T\ln\left[\prod_{k\in\partial i\backslash j}(Z_{k\rightarrow i}^{s}+Z_{k\rightarrow i}^{c})+e^{\beta\gamma}
        \prod_{k\in\partial i\backslash j}Z_{k\rightarrow i}^{s}\right]
\end{equation}
where $\beta\equiv1/T$ denotes the inverse temperature and the first
term $u^{2}/2$ is the energy of the reference state ({\em s} state).
$Z_{k\rightarrow i}^{s}$ and $Z_{k\rightarrow i}^{c}$ are the cavity
partition functions for node $k$ taking {\em s} and {\em c} states
respectively. Note that the first term in the square bracket of
Eq.~(\ref{cavfe01}) corresponds to the case of node $i$ taking {\em
s} state without node $j$, while the second term the case of node
$i$ taking {\em c} state as its cavity state. The total partition
function is the sum of these two terms~\cite{Mezard-04}. Subtracting
the free energy before adding node $i$, i.e.,
$-T\ln\left[\prod_{k\in\partial i\backslash j}(Z_{k\rightarrow
i}^{s}+Z_{k\rightarrow i}^{c})\right]$, we get the free energy
change on adding node $i$, given by
\begin{equation}\label{cavfe02}
        \Delta F_{i\rightarrow j}=\frac{u^{2}}{2}-T\ln\left[1+e^{\beta\gamma}
        \prod_{k\in\partial i\backslash j}\psi_{k\rightarrow i}^{s}\right]
\end{equation}
where the cavity probability is given by
\begin{equation}\label{cavfe03}
        \psi_{i\rightarrow j}^{s}\equiv\frac{Z_{i\rightarrow j}^{s}}{Z_{i\rightarrow j}^{s}+Z_{i\rightarrow j}^{c}}=
        \frac{1}{1+e^{\beta\gamma}
        \prod_{k\in\partial i\backslash j}\psi_{k\rightarrow
        i}^{s}}.
\end{equation}

\begin{center}
\begin{figure}
         \includegraphics[bb=88 350 472 682,scale=0.8]{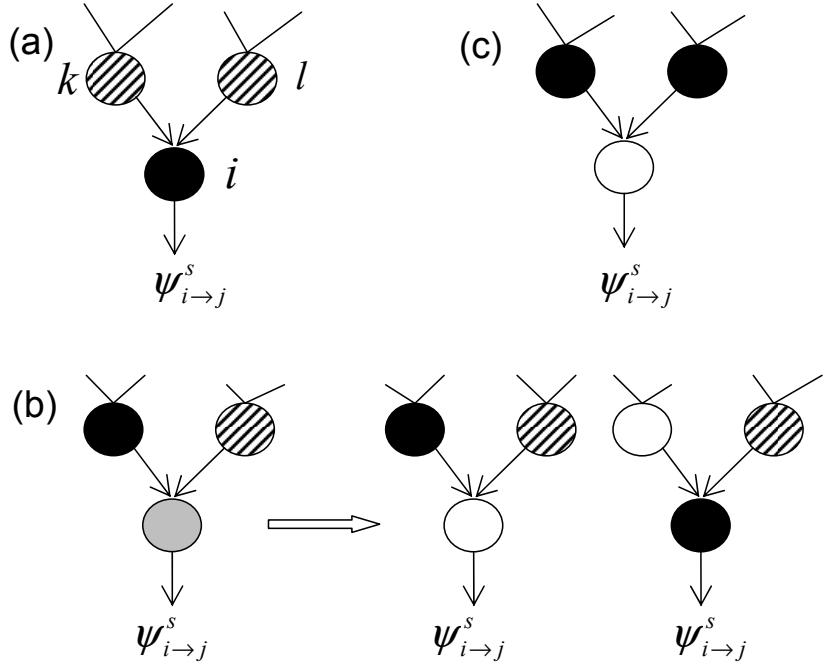}
  \caption{
             Entropic contribution due to node addition (adding node $i$ and its adjacent links). Each node is connected to $C=3$ nodes.
             The codes for the cavity probabilities are: shaded for $\psi_{k\rightarrow
             i}^{s}\in(0,1]$, black for $\psi_{k\rightarrow
             i}^{s}=0$, white for $\psi_{k\rightarrow
             i}^{s}=1$ and gray for $\psi_{k\rightarrow
             i}^{s}\in(0,1)$. The arrow shows the message passing direction. (a) None of the neighbors of node $i$ is
             frozen to the consumer state in the absence of node $i$. (b) Only one of the
             neighbors of node $i$, say node $k$, is frozen to
             the consumer state without node $i$. After the addition of node $i$, it can take the source state
             or consumer state. (c) At least two of neighbors of node $i$ are
             frozen to the consumer state in the absence of node $i$.
   }\label{site}
 \end{figure}
 \end{center}

The value for the cavity probability $\psi_{i\rightarrow j}^{s}$ of
node $i$ depends on the incoming cavity probabilities from its
neighbors other than node $j$, which can be categorized into three
cases.

 In the first case as depicted in
Fig.~\ref{site} (a), all neighbors of node $i$ (other than node $j$)
have non-zero cavity probabilities $\{\psi_{k\rightarrow i}^{s}\}$.
In this case, the cavity state of node $i$ must be a consumer in the
zero temperature limit, and $\Delta F_{i\rightarrow
j}=u^{2}/2-\gamma-T\ln\left[
        \prod_{k\in\partial i\backslash j}\psi_{k\rightarrow
        i}^{s}\right]$. Hence,
        \begin{subequations}\label{cavfe04}
        \begin{align}
          \psi_{i\rightarrow j}^{s}&=e^{-\beta\gamma-\Delta S_{i\rightarrow
        j}},\label{cavfe04a}\\
        \Delta S_{i\rightarrow
        j}&=\sum_{k\in\partial i\backslash
        j}\ln\psi_{k\rightarrow i}^{s},
        \end{align}
\end{subequations}
where $\Delta S_{i\rightarrow
        j}$ is the cavity entropy change when node $i$ and its adjacent edges except $(ij)$ are added (but the forward link $i\rightarrow j$ is considered in calculating the optimal state).

 In the second case (Fig.~\ref{site} (b)), only one neighbor
of node $i$, say node $k$, is frozen to the {\em c} state in the
zero temperature limit in the absence of node $i$, i.e.,
$\psi_{k\rightarrow i}^{s}$ is given by Eq.~(\ref{cavfe04a}). Then
we have $\Delta F_{i\rightarrow j}=\frac{u^{2}}{2}-T\ln\left[1+
        e^{-\Delta S_{k\rightarrow
        i}}\prod_{l\in\partial i\backslash k,j}\psi_{l\rightarrow
        i}^{s}\right]$, and
\begin{subequations}\label{cavfe05}
        \begin{align}
          \psi_{i\rightarrow
        j}^{s}&=\frac{1}{1+e^{-\Delta S_{k\rightarrow i}}\prod_{l\in\partial
i\backslash k,j}\psi_{l\rightarrow i}^{s}},\label{cavfe05a}\\
       \Delta S_{i\rightarrow
        j}&=\ln\left[1+e^{-\Delta S_{k\rightarrow i}}\prod_{l\in\partial i\backslash
k,j}\psi_{l\rightarrow i}^{s}\right].
        \end{align}
\end{subequations}

The third case where at least two of incoming $\psi_{k\rightarrow
i}^{s}$ for node $i$ vanish in the zero temperature limit is
presented in Fig.~\ref{site} (c). Based on Eqs.~(\ref{cavfe02})
and~(\ref{cavfe03}), we have
\begin{subequations}\label{cavfe06}
        \begin{align}
          \psi_{i\rightarrow j}^{s}&=1,\label{cavfe06a}\\
        \Delta S_{i\rightarrow
        j}&=0.
        \end{align}
\end{subequations}

In the above analysis, the added node is assumed to be deficient
node. However, a finite fraction of surplus nodes with very large
capacities are present in the transportation network as the quenched
disorder. The addition of a surplus node is assumed to have no
entropy contribution to the network and its full and cavity
probabilities are always fixed to be $1$ since it is frozen to the
source state by definition. Adding a surplus node will yield
different cavity energies depending on the states of its neighbors
since the consumer neighbors will draw resources from its adjacent
surplus nodes. This leads to the recursive relations
$\psi_{i\rightarrow
j}^{s}=\delta_{\Lambda_{i},A}+\delta_{\Lambda_{i},-1}\psi_{i\rightarrow
j}^{s}|_{\Lambda_{i}=-1}$ and $\Delta S_{i\rightarrow
j}=\delta_{\Lambda_{i},-1}\Delta S_{i\rightarrow
j}|_{\Lambda_{i}=-1}$, where $\psi_{i\rightarrow
j}^{s}|_{\Lambda_{i}=-1}$ and $\Delta S_{i\rightarrow
j}|_{\Lambda_{i}=-1}$ are given by the relevant expressions in
Eqs.~(\ref{cavfe04}) to~(\ref{cavfe06}). Note that, if we neglect
the entropic effects on the bistable nodes, the above analysis leads
to the belief propagation $\psi_{i\rightarrow
j}^{s}=\delta_{\Lambda_{i},A}+\delta_{\Lambda_{i},-1}\left[1-\prod_{k\in\partial
i\backslash j}\psi_{k\rightarrow i}^{s}\right]$ derived in
Ref.~\cite{Wong-2010}. Algorithmically, a decimation procedure
inspired by the fixed point solution of belief propagation can be
devised. We will compare this inspired decimation with the entropic
message passing algorithm in Sec.~\ref{sec_result}.
\begin{figure}
         \includegraphics[bb=78 398 517 726,scale=0.8]{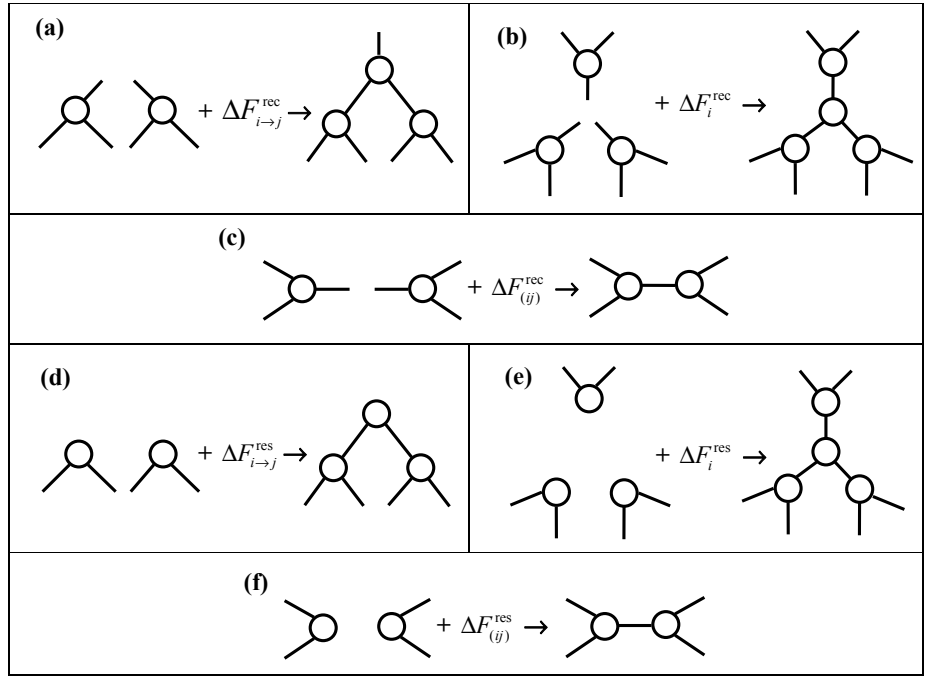}
  \caption{ Sketch for {\em
  reconnection} (a-c) and {\em restoration} (d-f) in the cavity iterations. (a) The recursion of
  the link-headed cavity free energy change given by Eq.~(\ref{cavfe02}).
  (b) The free energy change when node $i$ is reconnected, given by Eq.~(\ref{rec02}).
  (c) The free energy change when link $(ij)$ is reconnected, given
  by Eq.~(\ref{rec03}). (d) The recursion of the node-headed cavity
  free energy change, akin to a restoration process, given by Eq.~(\ref{restFe}). (e) The free energy change when node $i$ and its
  links are restored, given by Eq.~(\ref{rest03}). (f) The free
  energy change when link $(ij)$ is restored, given by
  Eq.~(\ref{rest06}).
   }\label{resrec}
 \end{figure}

\begin{figure}
         \includegraphics[bb=88 600 467 703,scale=0.8]{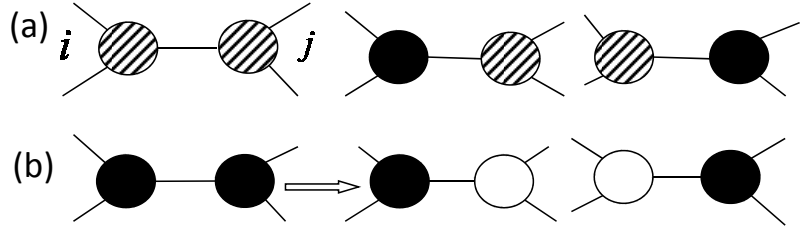}
  \caption{
         Entropic contribution due to edge addition. The
         probabilistic
         meaning for each node is the same as that in Fig.~\ref{site}. (a)
         At most one end of the added link is frozen
         into consumer state before the link addition. (b) Both ends of the added link
         are frozen into the consumer state in the absence of the link. Then either node $i$ or $j$ should change
         its state after the link is added.
   }\label{edge}
 \end{figure}

\subsubsection{Disconnecting and reconnecting a node and a link}
\label{RS03}

For a network with $N$ nodes and $L$ links, we consider an initial
configuration with $N-1$ nodes and $L$ links, obtained by
disconnecting node $i$ from its neighbors, while keeping the links
feeding node $i$ dangling in the network. In the dangling links, the
flow $x_{ik}$ is considered in optimizing the cavity energy of node
$k$ for all $k\in\partial i$. This allows node $k$ to take both the
{\em s} and {\em c} states. Hence the initial free energy is given
by
\begin{equation}\label{rec01}
    F_{N-1,L}=-T\sum_{k\in\partial i}\ln\left(Z_{k\rightarrow i}^{s}+Z_{k\rightarrow i}^{c}\right).
\end{equation}
Then we consider the final free energy after the node $i$ is
reconnected to its neighbors as shown in Fig.~\ref{resrec} (b).
Extending Eq.~(\ref{cavfe01}) to include all neighbors of node $i$,
we have
\begin{equation}\label{rec}
    F_{N,L}=\frac{u^{2}}{2}-T\ln\left[\prod_{k\in\partial i}(Z_{k\rightarrow i}^{s}+Z_{k\rightarrow i}^{c})+e^{\beta\gamma}
        \prod_{k\in\partial i}Z_{k\rightarrow i}^{s}\right].
\end{equation}
Thus the free energy change on reconnecting node $i$ is given by
\begin{equation}\label{rec02}
    \Delta F_{i}^{{\rm rec}}=\frac{u^{2}}{2}-T\ln\left[1+e^{\beta\gamma}
        \prod_{k\in\partial i}\psi_{k\rightarrow
        i}^{s}\right].
\end{equation}
Equation~(\ref{rec02}) is derived by subtracting Eq.~(\ref{rec01})
from Eq.~(\ref{rec}) and using the definition of
Eq.~(\ref{cavfe03}). The entropy change $\Delta S_{i}^{\rm rec}$ of
reconnecting node $i$ can then be computed in the zero temperature
limit as
\begin{equation}\label{rec05}
\begin{split}
\Delta S_{i}^{{\rm rec}}&=\Theta\left(\prod_{k\in\partial
i}\psi_{k\rightarrow i}^{s}\right)\sum_{k\in\partial
i}\ln\psi_{k\rightarrow i}^{s}+
      \sum_{k\in\partial i}\left[1-\Theta(\psi_{k\rightarrow
i}^{s})\right]\Theta\left(\prod_{l\in\partial i\backslash
k}\psi_{l\rightarrow
      i}^{s}\right)\\
      &\times\ln\left[1+e^{-\Delta S_{k\rightarrow i}}\prod_{l\in\partial i\backslash
k}\psi_{l\rightarrow i}^{s}\right].
      \end{split}
      \end{equation}

The cavity free energy change akin to a reconnection process can be
defined according to the link-headed diagrams in Fig.~\ref{resrec}
(a). In this case, the cavity free energy change reduces to
Eq.~(\ref{cavfe02}).

To obtain the entropy contribution of an edge, we consider an
initial configuration with $N$ nodes and $L+1$ links, obtained by
breaking the link between nodes $i$ and $j$ to form two dangling
links, one from node $i$ and the other from node $j$. In the
dangling link from node $i$ (node $j$), the flow $x_{ji}$ ($x_{ij}$)
is considered in optimizing the cavity energy of node $i$ (node
$j$), independent of the cavity energy of node $j$ (node $i$). This
allows nodes $i$ and $j$ to take both {\em s} and {\em c} states.
Hence the initial free energy is given by $F_{N,L+1}=F_{i\rightarrow
j}+F_{j\rightarrow i}$ where $F_{i\rightarrow j}$ or
$F_{j\rightarrow i}$ is given by Eq.~(\ref{cavfe01}).

Now we consider the final free energy after reconnecting the link
between nodes $i$ and $j$ as shown in Fig.~\ref{resrec} (c). It is
more convenient to analyze the free energy change starting from the
network with $N-2$ nodes obtained by excluding nodes $i$ and $j$ and
all their adjacent links. This includes the following three cases.
(a) In both $\partial i\backslash j$ and $\partial j\backslash
 i$, there are one or more nodes in the {\em c} state, then both nodes $i$
 and $j$ will be in the {\em s} state. No flow is present on the link
 $(ij)$, and energy change $E_{N}-E_{N-2}=u^{2}$. (b) In either $\partial i\backslash j$ or $\partial j\backslash
 i$, there are one or more nodes in the {\em c} state, and all nodes in
 the other set are in the {\em s} state. In this case, nodes $i$ and
 $j$ will be in the {\em s} and {\em c} states ({\em c} and {\em s} states) respectively.
 There is a flow from $i$ to $j$ (from $j$ to $i$), and
 $E_{N}-E_{N-2}=u^{2}-\gamma$. (c) All nodes in $\partial i\backslash j$ and $\partial j\backslash
 i$ are in the {\em s} state. In this case, $i$ and $j$ will be either in
 the {\em s} and {\em c} states, or the {\em c} and {\em s} states respectively.
 Correspondingly, there is a flow from $i$ to $j$, or from $j$ to
 $i$, and $E_{N}-E_{N-2}=u^{2}-\gamma$. These three different cases lead
 to the following free energy change
\begin{equation}\label{rec03}
    \begin{split}
      \Delta F_{(ij)}^{{\rm rec}}&=-T\ln\Biggl[\left(1-\prod_{k\in\partial i\backslash
j}\psi_{k\rightarrow i}^{s}\right)\left(1-\prod_{l\in\partial
j\backslash i}\psi_{l\rightarrow
j}^{s}\right)+e^{\beta\gamma}\left(1-\prod_{k\in\partial i\backslash
j}\psi_{k\rightarrow i}^{s}\right)\prod_{l\in\partial j\backslash
i}\psi_{l\rightarrow
j}^{s}\\
+&e^{\beta\gamma}\left(1-\prod_{l\in\partial j\backslash
i}\psi_{l\rightarrow j}^{s}\right)\prod_{k\in\partial i\backslash
j}\psi_{k\rightarrow i}^{s}+2e^{\beta\gamma}\prod_{l\in\partial
j\backslash i}\psi_{l\rightarrow j}^{s}\prod_{k\in\partial
i\backslash j}\psi_{k\rightarrow i}^{s}\Biggl]\\
&+T\ln\left[1+e^{\beta\gamma}
        \prod_{k\in\partial i\backslash j}\psi_{k\rightarrow
        i}^{s}\right]+T\ln\left[1+e^{\beta\gamma}
        \prod_{l\in\partial j\backslash i}\psi_{l\rightarrow
        j}^{s}\right].
      \end{split}
\end{equation}
We have used Eq.~(\ref{cavfe01}) and the definition of the cavity
probability $\psi_{i\rightarrow j}^{s}$ in Eq.~(\ref{cavfe03}) to
derive Eq.~(\ref{rec03}). Taking the zero temperature limit, we
obtain $\Delta S_{\left(ij\right)}=\ln\left[\psi_{i\rightarrow
j}^{s}+\psi_{j\rightarrow
 i}^{s}-\psi_{i\rightarrow
j}^{s}\psi_{j\rightarrow
 i}^{s}\right]$ in cases (a) and (b), which is the entropy change in Fig.~\ref{edge}(a) and $\Delta S_{\left(ij\right)}=\ln\left[e^{-\Delta S_{i\rightarrow j}}+e^{-\Delta S_{j\rightarrow
 i}}\right]$ in case (c), corresponding to the entropy change in Fig.~\ref{edge} (b). To sum up, the entropy change due to the link
 reconnection is written as
\begin{equation}\label{rec04}
\begin{split}
        \Delta S_{\left(ij\right)}^{{\rm rec}}&=\Theta(\psi_{i\rightarrow j}^{s}+\psi_{j\rightarrow
        i}^{s})\ln\left[1-(1-\psi_{i\rightarrow j}^{s})(1-\psi_{j\rightarrow
 i}^{s})\right]\\
 &+\left[1-\Theta(\psi_{i\rightarrow j}^{s}+\psi_{j\rightarrow
        i}^{s})\right]\ln\left[e^{-\Delta S_{i\rightarrow j}}+e^{-\Delta S_{j\rightarrow
 i}}\right].
 \end{split}
\end{equation}

Since the energy in the source location problem is distributed among
the installation costs of the nodes and the transportation costs of
the links, it is possible to formulate an alternative cavity
analysis in which the cavity trees are terminated in nodes instead
of links. In this case, changes in the free energy and entropy of a
network can be obtained by considering a network with a node or a
link first removed and then restored. We call this a {\em
restoration} process, in contrast to the {\em reconnection} process
described in this section. Recursion relations and the processes of
restoring a node and a link are shown in Fig.~\ref{resrec} (d) to
(f) respectively. As derived in Appendix~\ref{app:RS02}, subtle
differences exist between the two processes, but both processes
yield the same results when physical quantities such as the entropy
per node are calculated.
\subsubsection{Entropy per node}
\label{RS04}

The entropy density of source location problem can be evaluated in
the Bethe approximation~\cite{MM-2009} through
\begin{equation}\label{Entropy}
        s=\left<\Delta S_{i}\right>-\frac{C}{2}\left<\Delta
        S_{\left(ij\right)}\right>
\end{equation}
where $\left<\cdot\right>$ denotes both the disorder average and the
average over the cavity message distribution. The entropy density
can be obtained by substituting into Eq.~(\ref{Entropy}) the entropy
changes on disconnecting and reconnecting a node and a link
(Eqs.~(\ref{rec05}) and~(\ref{rec04})). Remarkably, the result is
the same as that obtained by using the zero temperature limit of
Eqs.~(\ref{rest03}) and~(\ref{rest06}) (removing and restoring a
node and a link). This can be seen by noting that the additional
terms appearing in Eqs.~(\ref{rest04}) and~(\ref{rest07}) cancel
each other in Eq.~(\ref{Entropy}).

We evaluate the entropy density by population dynamics
algorithm~\cite{cavity-2001}. A population of $\mathcal {N}$ pairs
of $(\psi_{i\rightarrow j}^{s},\Delta S_{i\rightarrow j})$ is used
to approximate the joint distribution $P(\psi_{i\rightarrow
j}^{s},\Delta S_{i\rightarrow j})$ and its components are uniformly
updated by the new computed ones according to Eqs.~(\ref{cavfe04})
to~(\ref{cavfe06}). Usually, a number of iterations $\mathcal
{T}-\mathcal {T}_{0}$ are used to compute the entropy value with
$\mathcal {T}_{0}$ iterations for equilibration. Note that
Eq.~(\ref{Entropy}) gives a self-averaging entropy value in the
thermodynamic limit in the sense that the typical value of the
entropy computed by the population dynamics algorithm should be
consistent with that computed on single large
networks~\cite{MM-2009}.

\subsection{One-step replica symmetry breaking analysis}
\label{sec:RSB} When replica symmetry is broken, the single ground
state would break up into exponentially many ground states plus an
even larger set of metastable states acting as the dynamical traps
for any greedy search algorithm. In this case, one should take into
account the reshuffling of free energies of different states when
cavity iterations are performed, therefore, we write the replicated
free energy $\Phi(y)$~\cite{Monasson-1995prl} as
\begin{equation}\label{repf}
        e^{-yN\Phi(y)}\equiv\sum_{\alpha}e^{-yNf_{\alpha}}=\int dfe^{N(\Sigma(f)-yf)}
\end{equation}
where $\alpha$ indicates each state and $\Sigma(f)$ is the
complexity function counting states with given free energy density
$f$. A saddle point analysis of Eq.~(\ref{repf}) gives
$\Phi(y)=f^{*}-\Sigma(f^{*})/y$ where $f^{*}$ is determined by
$y=d\Sigma(f)/df$. The inverse pseudotemperature $y$ allows us to
weight differently the various states according to their free energy
densities while the usual inverse temperature $\beta=1/T$ selects
the energy of equilibrium configurations. Actually, Eq.~(\ref{repf})
corresponds to a decomposition of the Gibbs
measure~\cite{Montanari-2008jstat}. Here, our analysis is restricted
to the ground state ($\beta$ tends to infinity), and $f$ tends to
the energy density $\epsilon$ in this limit. For the current
problem, the energetic complexity $\Sigma(\epsilon)$ can be computed
by the following Legendre transform
\begin{subequations}\label{complexity}
\begin{align}
\Sigma(\epsilon)&=y(\epsilon-\Phi),\\
\epsilon&=\frac{\partial(y\Phi)}{\partial y},
\end{align}
\end{subequations}
where $\Sigma(\epsilon)$ can be computed from the parametric plot of
$\epsilon(y)$ and $\Sigma(y)$ by varying the value of $y$. We remark
here that the complexity function of the source location problem
vanishes at finite $y$~\cite{Wong-2010} and this was also observed
in studies of the minimal vertex cover problem~\cite{ZH-pre06}.
Therefore, the Parisi replica symmetry breaking parameter defined by
$m=y/\beta$ vanishes in the zero temperature limit. The parameter
$m$ is used to select the size of the investigated ground state
(entropy), as studied in
Refs.~\cite{Mezard-2005prl,Lenka-2007pre,Montanari-2008jstat} to
compute the entropic complexity curve for random constraint
satisfaction problems in the zero ground state energy region and the
replicated free energy $\Phi$ reaches the maximum at $y=\infty$(in
this case, $m\in[0,1]$~\cite{ZH-2008}).

Due to the proliferation of pure states, we write the recursive
equations for the joint distribution $P_{i\rightarrow
j}(\psi_{i\rightarrow j}^{s},\Delta S_{i\rightarrow j})$ at the
$1$RSB level as
\begin{equation}\label{EcmRSB}
\begin{split}
        P_{i\rightarrow j}(\psi_{i\rightarrow
j}^{s},\Delta S_{i\rightarrow j})=\mathcal {Z}_{i\rightarrow
j}^{-1}\int\prod_{k\in\partial i\backslash j}dP_{k\rightarrow
i}(\psi_{k\rightarrow i}^{s},\Delta S_{k\rightarrow i})e^{-y\Delta
E_{i\rightarrow j}}\\
\times\delta\left[\psi_{i\rightarrow j}^{s}-\mathcal
{F}(\{\psi_{k\rightarrow i}^{s},\Delta S_{k\rightarrow
i}\})\right]\delta\left[\Delta S_{i\rightarrow j}-\tilde{\mathcal
{F}}(\{\psi_{k\rightarrow i}^{s},\Delta S_{k\rightarrow i}\})\right]
 \end{split}
\end{equation}
where $dP_{k\rightarrow i}(\psi_{k\rightarrow i}^{s},\Delta
S_{k\rightarrow i})\equiv d\psi_{k\rightarrow i}^{s}d\Delta
S_{k\rightarrow i}P_{k\rightarrow i}(\psi_{k\rightarrow
i}^{s},\Delta S_{k\rightarrow i})$. The functions $\mathcal {F}$ and
$\tilde{\mathcal {F}}$ are given by the relevant expressions of
$\psi_{i\rightarrow j}^{s}$ and $\Delta S_{i\rightarrow j}$ in
Eqs.~(\ref{cavfe04}) to~(\ref{cavfe06}). The reweighting factor
$e^{-y\Delta E_{i\rightarrow j}}$ takes into account the energy
change due to cavity operation for the reconnection process (the
addition of node $i$, the dangling edge $(ij)$ and the edges to its
neighbors other than $j$). Note that $m=0$ makes the contribution of
the entropy change in this term disappear. The cavity energy values
of $\Delta E_{i\rightarrow j}$ are obtained by the zero temperature
limit of Eq.~(\ref{cavfe02}), yielding $u^{2}/2-\gamma$ if none of
the neighbors $k\in\partial i\backslash j$ takes {\em c} state, and
$u^{2}/2$ otherwise. Finally, $\mathcal {Z}_{i\rightarrow j}$ is a
normalization constant. To simplify the analysis, we parameterize
the joint distribution according to the discussion in
Sec.~\ref{sec:RS} as
\begin{equation}\label{param}
\begin{split}
       P_{i\rightarrow j}(\psi_{i\rightarrow
j}^{s},\Delta S_{i\rightarrow j})&=p_{i\rightarrow
j}^{1}\delta(\psi_{i\rightarrow j}^{s}-1)\delta(\Delta
S_{i\rightarrow j})+p_{i\rightarrow j}^{0}\delta(\psi_{i\rightarrow
j}^{s})\tilde{\phi}^{0}_{i\rightarrow j}(\Delta S_{i\rightarrow
j})\\
&+p_{i\rightarrow j}^{*}\tilde{\phi}^{*}_{i\rightarrow
j}(\psi_{i\rightarrow j}^{s},\Delta S_{i\rightarrow j})
\end{split}
\end{equation}
where $\int d\tilde{\phi}^{0}_{i\rightarrow j}(\Delta
S_{i\rightarrow j})=1$, $\int d\tilde{\phi}^{*}_{i\rightarrow
j}(\psi_{i\rightarrow j}^{s},\Delta S_{i\rightarrow j})=1$ and
$p_{i\rightarrow j}^{1}+p_{i\rightarrow j}^{0}+p_{i\rightarrow
j}^{*}=1$. Compared with the RS case where the messages merely
consist of the pair $(\psi_{i\rightarrow j}^{s},\Delta
S_{i\rightarrow j})$ for each directed edge, here the order
parameter turns out to be a survey of these messages at the $1$RSB
level; $p_{i\rightarrow j}^{1}$, $p_{i\rightarrow j}^{0}$ and
$p_{i\rightarrow j}^{*}$ tell us the probability of picking up a
pure state at random and finding that the cavity state of node $i$
is source, consumer and free respectively. These three surveys
enable us to handle the state-to-state fluctuations at the $1$RSB
level. With this parametric representation, a finite $y$ survey
propagation equation can be obtained for the source location problem
as
\begin{subequations}\label{SPy}
\begin{align}
  p_{i\rightarrow j}^{0}&=\delta_{\Lambda_{i},-1}\frac{e^{y\gamma}\prod_{k\in\partial i\backslash j}(p_{k\rightarrow i}^{1}+p_{k\rightarrow i}^{*})}
  {1+(e^{y\gamma}-1)\prod_{k\in\partial i\backslash j}(p_{k\rightarrow i}^{1}+p_{k\rightarrow
  i}^{*})},\\
  p_{i\rightarrow j}^{*}&=\delta_{\Lambda_{i},-1}\frac{\sum_{k\in\partial i\backslash j}p_{k\rightarrow i}^{0}\prod_{l\in\partial i\backslash j,k}
  \left(p_{l\rightarrow i}^{1}+p_{l\rightarrow i}^{*}\right)}{1+(e^{y\gamma}-1)\prod_{k\in\partial i\backslash j}(p_{k\rightarrow i}^{1}+p_{k\rightarrow
  i}^{*})},\\
  p_{i\rightarrow
  j}^{1}&=\delta_{\Lambda_{i},A}+\delta_{\Lambda_{i},-1}(1-p_{i\rightarrow j}^{0}-p_{i\rightarrow
  j}^{*}).
\end{align}
\end{subequations}
After the fixed point of Eq.~(\ref{SPy}) is obtained, the replicated
free energy can be computed via
\begin{equation}\label{repf02}
     \Phi(y)=\left<\Delta\Phi_{i}\right>-\frac{C}{2}\left<\Delta\Phi_{\left(ij\right)}\right>
\end{equation}
where the average is taken over the capacity distribution and the
survey distribution, and the replicated free energy shift
$\Delta\Phi_{i}$ due to node addition (and its $C$ edges) and
$\Delta\Phi_{\left(ij\right)}$ due to link addition are,
respectively,
\begin{subequations}\label{cavrpf}
\begin{align}
  -y\Delta\Phi_{i}&=\delta_{\Lambda_{i},-1}\left\{-y\frac{u^{2}}{2}+\ln\left[1+(e^{y\gamma}-1)\prod_{k\in\partial i}\left(p_{k\rightarrow i}^{1}+p_{k\rightarrow
  i}^{*}\right)\right]\right\},\label{cavrpfa}\\
  -y\Delta\Phi_{\left(ij\right)}&=\ln\Biggl[1-(1-e^{-y\gamma})p_{i\rightarrow j}^{0}p_{j\rightarrow i}^{0}\Biggr]\label{cavrpfb},
\end{align}
\end{subequations}
where we have used the energy changes calculated in the zero
temperature limits of Eqs.~(\ref{rec02}) and~(\ref{rec03}), rather
than those of Eqs.~(\ref{rest03}) and~(\ref{rest06}). In
Eq.~(\ref{cavrpfa}), we consider the energy change when node $i$ is
reconnected to the network, instead of the change when node $i$ and
its links to neighbors are restored to the network. This is because
the dangling forward links $i\rightarrow j$ have already been
included in the recursion relations of $p_{i\rightarrow j}^{1}$,
$p_{i\rightarrow j}^{0}$ and $p_{i\rightarrow j}^{*}$. Similarly in
Eq.~(\ref{cavrpfb}), we consider the energy change when the dangling
forward links $i\rightarrow j$ and $j\rightarrow i$ are reconnected
to form the link $(ij)$, instead of the change when link $(ij)$ is
restored to the network, since the dangling forward links have
already been included in the recursion relations of $p_{i\rightarrow
j}^{1}$, $p_{i\rightarrow j}^{0}$ and $p_{i\rightarrow j}^{*}$. The
steps involved in this procedure are shown schematically in
Fig.~\ref{resrec} (a)-(c).

The energetic complexity can then be computed using
Eq.~(\ref{complexity}). In the glassy phase, $\Sigma(\epsilon)$
typically increases from $y=0$ up to the maximal point forming the
first non-physical convex part, yet, with further increase in $y$,
decreases down to the zero point where the complexity vanishes at
the ground state energy ($y=y^{*}$). This second branch of the
complexity curve is the physical concave part~\cite{cavity-2003}.
Actually, the zero complexity corresponds to the maximum of the
replicated free energy since $\Sigma=y^{2}\partial_{y}\Phi$ from the
Legendre transform. To compute the ground state entropy at the
$1$RSB level, we should fix $y=y^{*}$.

To derive the formula for the ground state entropy at the $1$RSB
level, we first write the replicated free energy
$-y\Phi(y,m)=\Sigma(\epsilon,s)-y\epsilon+ms$ keeping a finite value
of $m$~\cite{Mezard-2005prl,ZH-2008} and at the end of the
derivation, we get the ground state entropy via
$s=\frac{\partial(-y\Phi(y,m))}{\partial m}|_{m=0}$ where $y=y^{*}$
determined by $\Sigma(y^{*})=0$. The $1$RSB approximation of ground
state entropy density reads,
\begin{subequations}\label{EntropyRSB}
\begin{align}
     s&=\left<\frac{\left[\Delta S_{i}e^{-y\Delta E_{i}}\right]}{\left[e^{-y\Delta
     E_{i}}\right]}\right>
     -\frac{C}{2}\left<\frac{\left[\Delta S_{\left(ij\right)}e^{-y\Delta E_{\left(ij\right)}}\right]}{\left[e^{-y\Delta E_{\left(ij\right)}}\right]}\right>,\label{S_RSB}\\
     \left[\Delta S_{i}e^{-y\Delta E_{i}}\right]&=\int\prod_{k\in\partial i}dP_{k\rightarrow
i}(\psi_{k\rightarrow i}^{s},\Delta S_{k\rightarrow i})e^{-y\Delta
E_{i}}\Delta S_{i}\left(\{\psi_{k\rightarrow i}^{s},\Delta
S_{k\rightarrow i}\}\right),\\
     \left[e^{-y\Delta E_{i}}\right]&=\delta_{\Lambda_{i},-1}e^{-yu^{2}/2}\left\{1+(e^{y\gamma}-1)\prod_{k\in\partial i}\left(p_{k\rightarrow i}^{1}+p_{k\rightarrow
  i}^{*}\right)\right\},\\
  \left[\Delta S_{\left(ij\right)}e^{-y\Delta E_{\left(ij\right)}}\right]&=\int dP_{i\rightarrow
j}(\psi_{i\rightarrow j}^{s},\Delta S_{i\rightarrow
j})dP_{j\rightarrow i}(\psi_{j\rightarrow i}^{s},\Delta
S_{j\rightarrow i})e^{-y\Delta E_{\left(ij\right)}}\Delta
S_{\left(ij\right)},\\
  \left[e^{-y\Delta E_{\left(ij\right)}}\right]&=1-(1-e^{-y\gamma})p_{i\rightarrow j}^{0}p_{j\rightarrow i}^{0}.
\end{align}
\end{subequations}
Note that in the average $[\cdot]$, $\Delta S_{i}$ and $\Delta
S_{\left(ij\right)}$ are still given by RS equations obtained from
the zero temperature limits of the free energies in
Eqs.~(\ref{rec02}) and~(\ref{rec03}) respectively, but the incoming
messages to compute them should be sampled from the joint
distribution Eq.~(\ref{param}), taking the reweighting factor into
account. In addition, $\left<\cdot\right>$ in the first term and
second term in Eq.~(\ref{S_RSB}) denote the averages over the
capacity distribution and survey distribution and this average can
be easily done by the population dynamics
algorithm~\cite{cavity-2001}, which yields a typical value of the
entropy density.

We remark that the expressions of $\Delta\Phi_{i}$ and
$\Delta\Phi_{(ij)}$ obtained in Ref.~\cite{Wong-2010} are different
from those in this paper. This is because in Ref.~\cite{Wong-2010},
the reweighting factors are based on the cavity free energies
following the node-headed recursions in the restoration process
schematically depicted in Fig.~\ref{resrec} (d). Hence when
$\Delta\Phi_{i}$ and $\Delta\Phi_{(ij)}$ are calculated,
Ref.~\cite{Wong-2010} used the free energy changes in
Fig.~\ref{resrec} (e) and (f) respectively. However, the finite $y$
survey propagation equation for both reconnection and restoration
cases can be verified to be equivalent through an algebraic
transformation.

\subsection{The appearance of frozen variables as an indicator of $1$RSB, and their distribution}

Another way to characterize the glassy behavior is to consider the
effects of freezing~\cite{Boet-2004,ZH-prl05}. We assume the set
$\mathcal {D}$ to be dynamical variables (other variables are
quenched sources, and treated only as a boundary condition). Frozen
variables in our case are those deficient nodes taking the same
state (source or consumer) in all ground states. It has been argued
that the RSB transition in vertex cover is related to the ability of
a quenched variable to induce a long range rearrangement of the
ground states~\cite{ZH-prl05}. In our problem there is a freezing
process due to the appearance of long-range correlations related to
RSB, but also due to a simple topological element: deficient nodes
close to many surplus nodes tend to be frozen simply due to this
proximity, independent of the rest of the network. Indeed it is much
easier to develop a theory of the latter effect than the former. For
this reason it is useful to introduce some (standard) granularity in
the network description when discussing freezing.

We consider a subgraph, where only connections amongst dynamical
variables are considered. On this subgraph there is a $2$-core, that
is the graph that remains after recursively removing dynamical
variables of connectivity one or lower. If the thermodynamics of the
$2$-core is simple so is the entire graph, variables outside the
$2$-core are in tree like structures that cannot exhibit a RSB
behavior independently of the $2$-core.

Let us consider the ensemble of regular $C=3$ random graphs with a
fraction ($1-\phi_s$) of deficient nodes, as later studied
numerically. We can further refine our definition of deficient nodes
on the $2$-core: those nodes of connectivity $3$ we call hubs, and
those of connectivity $2$ non-hubs. The simplest case is $\phi_s=0$,
in which the entire graph is the $2$-core, and all variables are
hubs (but this is the exception). By contrast when
$\phi_{s}>\frac{1}{C-1}$, the quenched nodes are so numerous that
the dynamical variables become disconnected in components of size
$o(N)$, that is the $2$-core disappears and there are no hubs. The
fraction of hubs in the $2$-core can be calculated analytically by
the following procedure. First, we define a probability $p_{dd}$
that a deficient node in the $2$-core is dangling (of connectivity
$1$), and this happens only when the states for $C-1$ of its
neighbors are either surplus or dangling deficient. Thus $p_{dd}$
satisfies the recursive equation
$p_{dd}=\left[\phi_{s}+(1-\phi_{s})p_{dd}\right]^{C-1}$, whose
stable fixed point is denoted as $p_{dd}^{*}
(=\phi_{s}^{2}/(1-\phi_{s})^{2}$ for $C=3$). A deficient node is a
hub only when all of its $C$ neighbors are neither surplus nor
dangling deficient, therefore the fraction of hubs in the $2$-core
is $ f_{{\rm
    hubs}}=(1-\phi_{s})\left[(1-\phi_{s})(1-p_{dd}^{*})\right]^{C}$. For $C=3$, we have
\begin{equation}\label{Anahubs}
    f_{{\rm
    hubs}}=\frac{(1-2\phi_{s})^{3}}{(1-\phi_{s})^{2}}.
\end{equation}

The hubs can be loosely considered as those nodes furthest from the
surplus nodes, but also control the phenomena of freezing. Let us
consider for example a system in which all hubs are quenched to
particular values (frozen and inflexible). Now if any non-hub is
quenched to a particular value, will it cause an extensive
rearrangement of the ground states? The freezing of the hubs
indicates that this propagation would be restricted to only a short
chain within the $2$-core, and terminate at the nearest hubs. This
indicates the phenomenological importance of the hubs. In order for
information from the quenched variable to propagate effectively,
there must exist a freedom of hubs to change state and thereby
branch the information outward. Of course frozen hubs can be
perturbed if we allow for a change in energy, the range of
propagation of information would then depend on the nature of the
freezing\textemdash if the freezing is caused by proximity to a
boundary, there will again be no long range reordering, whereas if
the freezing happens in a graph-wide correlated manner, there will
be a long range reordering of the system that is typical of RSB.

\subsection{Population dynamics}
\label{PD} We evaluate the entropy numerically by population
dynamics algorithm. At the $1$RSB level, we create a population of
$\mathcal {N}$ pairs of $(p_{i\rightarrow j}^{0},p_{i\rightarrow
j}^{*})$ with an additional population of $\mathcal {M}$ pairs of
$(\psi_{i\rightarrow j}^{s},\Delta S_{i\rightarrow j})$ associated
with each element of the first population. Elements of both
populations are updated in the population dynamics iterations. A
number of iterations $\mathcal {T}-\mathcal {T}_{0}$ are used to
compute the entropy value with $\mathcal {T}_{0}$ iterations for
equilibration. Another relevant quantity is the fraction of source
nodes $f_{s}$ in the final optimized network. Its typical value can
be evaluated as
$f_{s}=\frac{1}{N}\sum_{i=1}^{N}\left<\psi_{i}^{s}\right>$ at the RS
level or $1$RSB level. The average is taken over the disorder and
the RS message or $1$RSB survey distribution, and $\psi_{i}^{s}$ is
the full or marginal probability given by
$\psi_{i}^{s}=\delta_{\Lambda_{i},A}+\psi_{i}^{s}|_{\Lambda_{i}=-1}$
where $\psi_{i}^{s}|_{\Lambda_{i}=-1}$ is determined by
Eq.~(\ref{cavfe04a}),~(\ref{cavfe05a}) and~(\ref{cavfe06a}) in which
$\partial i\backslash j$ is replaced by $\partial i$.

\begin{figure}
\centering
    \includegraphics[bb=18 16 292 219,scale=1.0]{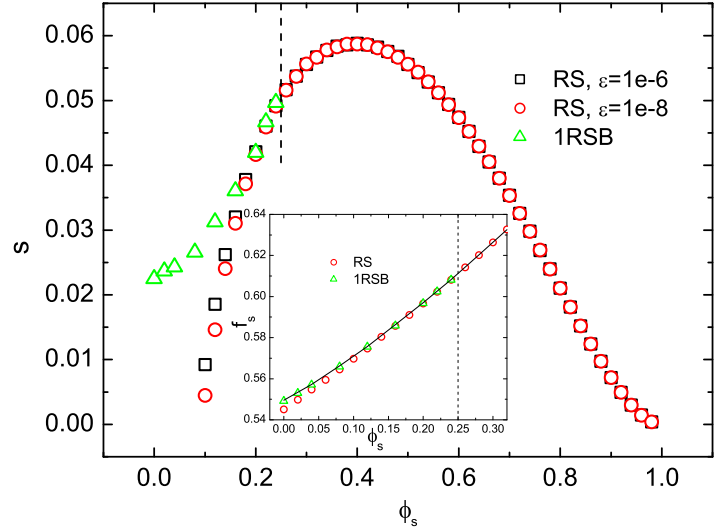}
  \caption{
  (Color online) Ground state entropy density versus the fraction of
  surplus nodes $\phi_{s}$. The vertical dashed line indicates the
  dynamical transition below which the RS description becomes incorrect.
  The RS results are obtained using different cutoffs $\varepsilon=10^{-6}$
  and $10^{-8}$. The network parameter $C=3$
  and the population size for RS computation is $\mathcal {N}=10^5$,
  and $(\mathcal {N},\mathcal {M})=(2048,1024)$ for $1$RSB
  computation. The inset gives the typical ground state fraction of source
  nodes in the final optimized network. The solid line is the asymptotic limit obtained by EO.
  }\label{GS}
\end{figure}

In the following studies, we choose $u=2/3$ and $C=3$ which fall
within the range of the singlet regime
$1/\sqrt{C}<u<\sqrt{(C+1)/[C(C-1)]}$~\cite{Wong-2010}. The entropy
values of the ground state at both RS and $1$RSB levels are given in
Fig.~\ref{GS}. We use $(\mathcal {T},\mathcal {T}_{0})=(5000,1000)$
for RS computation and $(\mathcal {T},\mathcal {T}_{0})=(1500,500)$
for $1$RSB computation. Only the mean value of the entropy density
is plotted since the error bar is much smaller than the symbol size.
In the RS population dynamics, the extremely small value for certain
cavity probabilities can be observed when $\phi_{s}$ is sufficiently
small, and this is equivalent to the divergence of the evanescent
cavity fields~\cite{Huang-pre09,Pan-2009}. However, if we take a
proper cutoff $\varepsilon$ for the nearly vanishing probability,
this situation can be circumvented~\cite{ZH-pre09} and the resulting
entropy value seems to be insensitive to the cutoff value as long as
$\varepsilon\geq10^{-6}$ in the stable region of RS solution. For
the $1$RSB computation, we adopt $\varepsilon=10^{-6}$ (smaller
values of $\varepsilon$ do not change much the result). In
Fig.~\ref{GS}, the entropy density first increases when $\phi_{s}$
decreases, then reaches a maximum followed by the decreasing trend
as the fraction of surplus nodes (disorder) further decreases. At a
large value of $\phi_{s}$, most nodes are surplus nodes, the entropy
should take a small value; when $\phi_{s}$ gets close to zero, most
nodes are deficient nodes, which makes the source locations much
more constrained and yields a small value of entropy. Due to the
concavity of the entropy function, there should exist some maximum
point and this point can be identified by the mean field computation
in Fig.~\ref{GS}. The maximum implies that at the corresponding
$\phi_{s}\simeq0.40$, there exists the largest number of optimal
assignments satisfying the singlet connection pattern among all
values of $\phi_{s}$.

At $\phi_{s}=0.25$, the replica symmetric solution starts to be
unstable against $1$RSB perturbations~\cite{Wong-2010}. This is
confirmed by the fact that above this transition point, the
complexity calculated by $1$RSB equations~(\ref{SPy}) does not
depend on the value of $y$ and is always zero, while it has a finite
value below the transition point. This transition point is also
called the dynamical transition threshold below which the local
search process is usually trapped by the metastable states. For the
source location problem, the replicated free energy reaches its
maximum at finite $y$ and actually the maximum corresponds to the
ground state~\cite{cavity-2003}. In the zero temperature limit, one
can thus obtain the ground state energy. Within the current context,
the ground state entropy value in the $1$RSB ansatz can also be
computed improving the RS prediction. We also show the fraction of
source nodes in the ground state in the inset of Fig.~\ref{GS} where
the $1$RSB ansatz predicts a higher value compared with the RS
ansatz. The $1$RSB result is consistent with the asymptotic limit
obtained by EO (see Sec.~\ref{NumEO}). We compute $f_{s}$ through
the linear relationship
$f_s=(\epsilon-1/(2C)+\phi_{s}u^{2}/2)/\gamma$ since we have the
identity $(1-f_{s})/(2C)+(f_{s}-\phi_{s})u^{2}/2=\epsilon$. In this
sense, the $f_{s}$ serves as a measure of the ground state energy
density. These values will be compared with those obtained from
simulations on single instances in Sec.~\ref{sec_result}. As shown
in the inset of Fig.~\ref{GS}, $f_{s}$ decreases as $\phi_{s}$
decreases. Actually, one deficient node causes an installation cost
($u^{2}/2$) when converted to a source node while it leads to an
increment of $1/(2C)$ of the transportation cost when remaining as a
consumer node. Notice that the consumer state is energetically
favored since $\gamma>0$ in the singlet regime. Consequently, on one
hand, as $\phi_{s}$ decreases, the number of deficient nodes
increases, on the other hand, to yield the minimal energy cost, some
deficient nodes need to remain as consumer nodes while consumer
nodes are not allowed to be paired in the singlet regime. The
competition of these two effects leads to the decreasing trend of
$f_{s}$ with decreasing $\phi_{s}$.

\begin{center}
\begin{figure}
    \includegraphics[bb=53 52 405 301,scale=1.0]{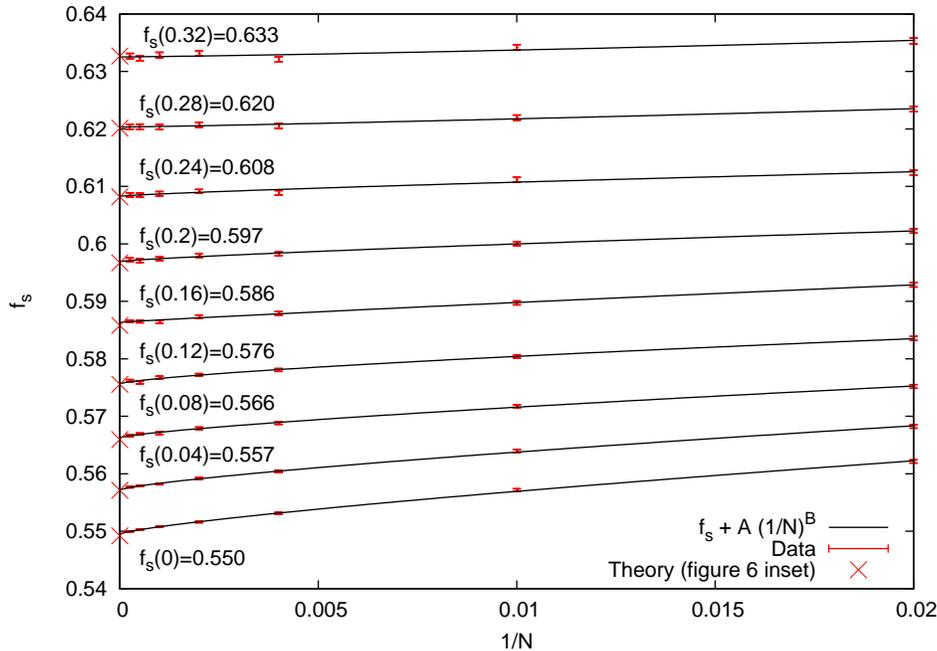}
  \caption{
  (Color online) Ground state energy achieved by EO. A power
  law fitting as $f_{s}(N,\phi_{s})=f_s(\phi_{s})+AN^{-B}$ is applied. For
  $\phi_{s}>0.25$, $B\simeq1.0$, however, $B$ decreases to a value
  $\simeq 0.8$ for $\phi_{s}$ close to zero.
  }\label{energy}
\end{figure}
\end{center}

\begin{center}
\begin{figure}
    \includegraphics[bb=53 52 403 298,scale=1.0]{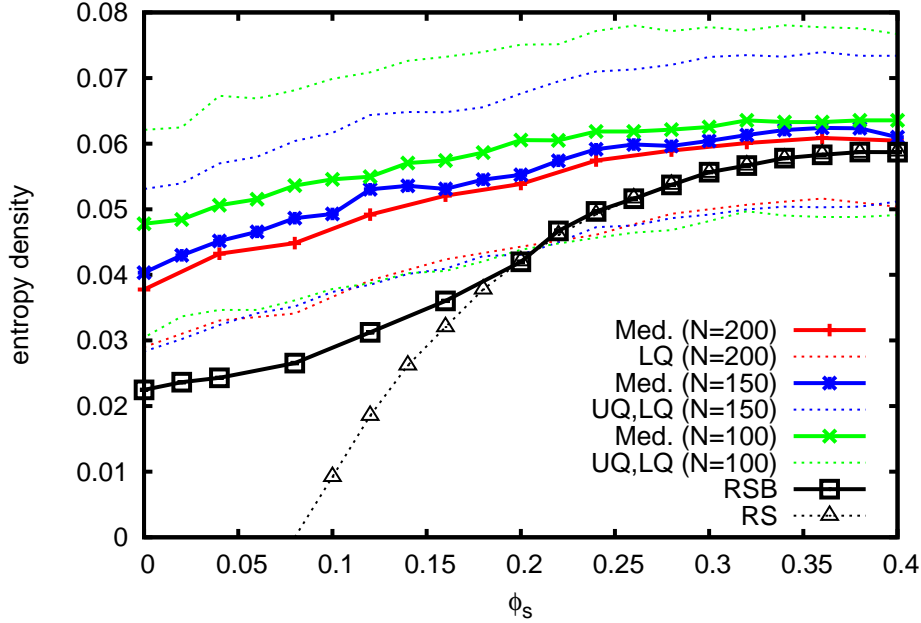}
  \caption{
  (Color online) The medians and upper and lower quartiles (Med, UQ and LQ) for
the entropy distribution for a set of $1500$, $1000$ and $500$
graphs for $N=100$, $150$ and $N=200$ respectively. For $N=200$ the
entropy becomes very large, meaning that only the median and lower
quartile can be established.
  }\label{entro}
\end{figure}
\end{center}

\subsection{Numerical verification by extremal optimization}
\label{NumEO} Extremal Optimization (EO) is a local stochastic
search method that has been employed successfully to understand the
ground-state properties of spin-glass like
systems~\cite{Boet-2003,Boet-2004,Raymond-2012}. It succeeds by
searching with a scale-free rule for changing states, in such a way
that it reaches low energy states, but is not trapped by large
energy barriers typical of systems that exhibit RSB behavior. In
this paper it is employed to experimentally verify the energy and
entropy of the singlet regime, as well as to investigate freezing
properties. The implementation details of this procedure are shown
in Appendix~\ref{sec:EO}. EO has been used to study the average
ground state energy, backbone, entropy and local field distribution
of spin glass
models~\cite{Boet-2003,Boet-2003epjb,Boet-2004,Boet-2008}. Our use
of EO in the exploration of extensive entropic properties is a new
application, but can be justified along the same lines as previous
studies: we know that EO will not sample uniformly the ground
states, however to establish the entropy we need only count the
ground states for which a biased sampler can be suitable as
explained in the Appendix~\ref{sec:EO} and justified empirically.

Data from various system sizes can be combined to estimate
asymptotic result and quantify finite size effects, as plotted in
Fig.~\ref{energy}. We have fitted a curve of the form
$f_{s}(N,\phi_{s})=f_s(\phi_{s})+AN^{-B}$, to our energy data, where
$f_s(\phi_{s})$ is our estimate of the asymptotic ground state
energy. Using the Marquardt-Levenberg method to fit the data, we
acquire the curves in Fig.~\ref{energy} in which the error bars
associated to the sample mean are plotted, from which the limiting
values are obtained with errors smaller than $10^{-3}$ in every
case. The exponent $B$ varies from approximately $0.785\pm 0.025$ at
$\phi_s=0$, to statistically indistinguishable from linear
($B\simeq1.0$) for $\phi_s\gtrsim 0.25$. The asymptotic limit
  $f_s(\phi_{s})$ shown in the plot is consistent with the
  theoretical prediction in Fig.~\ref{GS} (inset). Fig.~\ref{entro} shows the entropy obtained by EO as
described in Appendix~\ref{sec:EO}. The quartiles and medians of the
statistics indicate a broad range of entropies. For low $\phi_{s}$ ,
especially in the glassy region described by RSB, the median curves
of the EO statistics appear to approach a limiting curve from above.
However, the combination of lower quartile and median curves
indicates there may be a significant gap to the $1$RSB theory. Given
the network size less than $200$, this may be due to finite size
effects, or may be related to higher order RSB corrections. Since
the numerical method employed involves recording and comparing all
ground states, we are restricted to studying systems of size
$N\lesssim 200$. The data, in particular the lower quartile curve,
indicates a concentration of the entropy at an extensive value, in
qualitative agreement with the $1$RSB result for smaller $\phi_{s}$,
and in quantitative agreement with RS and $1$RSB for larger
$\phi_{s}$, which furthermore confirms that $1$RSB becomes a better
approximation than RS in the low $\phi_{s}$ regime.

\begin{figure}
\centering
    \includegraphics[bb=13 16 284 220,scale=1.0]{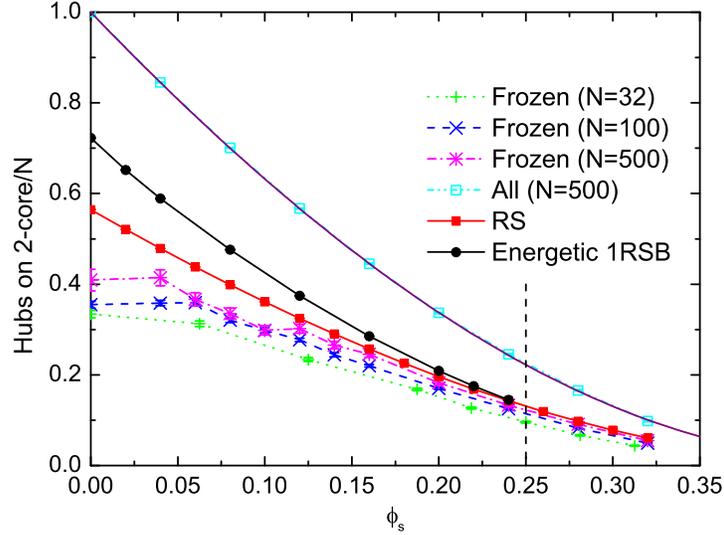}
  \caption{
  (Color online) Statistics of hubs on $2$-core versus the
  quenched disorder $\phi_{s}$. The upper solid line is the analytic
  prediction for the fraction of hubs on $2$-core. It predicts the
  numerical data on graphs of $N=500$ very well. The vertical dashed line indicates the
  dynamical transition point. The freezing phenomena is predicted well for $\phi_s\gtrsim 0.12$,
 at which point there is a deviation between the theory and numerical results.
  }\label{hubs}
\end{figure}

\begin{figure}
\centering
    \includegraphics[bb=10 13 301 227,scale=1.0]{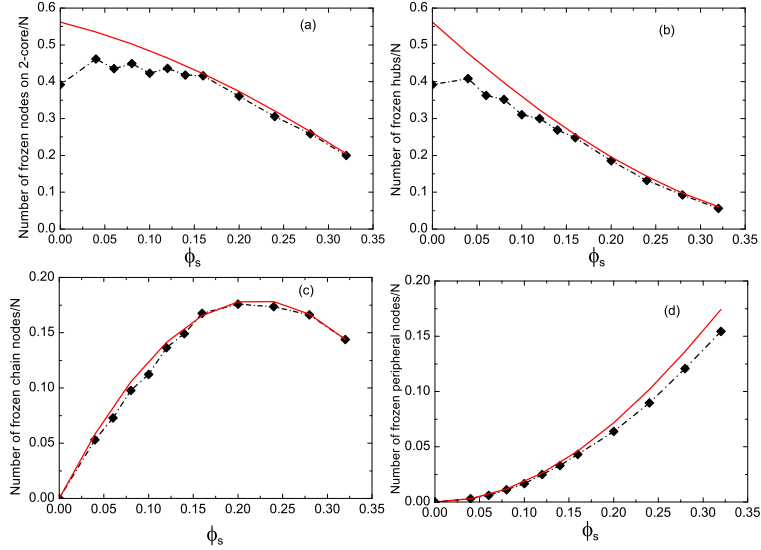}
  \caption{
  (Color online) The dependence of the fraction of frozen nodes on the fraction $\phi_{s}$ of surplus nodes. (a) All frozen $2$-core nodes considered. (b)
Frozen hubs. (c) Frozen chain nodes. (d) Frozen peripheral nodes.
Lines: RS prediction. Symbols: Results obtained by the EO algorithm,
extrapolated linearly to the infinite size limit (based on data at
$N=50, 100, 250, 500$ and $1000$ with $2500, 2500, 1000, 240$ and
$60$ samples respectively).
  }\label{substruct}
\end{figure}

The effect of freezing can also be probed with the EO numerical
method (see Appendix~\ref{sec:EO}). Fig.~\ref{hubs} shows the mean
density of hubs and density of frozen hubs (based on $2500,2500,240$
samples for graphs of $N=32,100,500$ respectively). We also plot the
entropic RS prediction and energetic $1$RSB prediction for the
frozen hubs in Fig.~\ref{hubs}. The numerical data for high $\phi_s$
is consistent with the theoretical prediction, while for small
$\phi_s$ the numerical results continue to evolve rapidly with $N$
towards the theoretical curves from below. We find that the $1$RSB
theory predicts a higher value of the fraction of frozen hubs after
the dynamical transition than RS. However, the theoretical
prediction of the fraction of frozen hubs is higher than that
measured in numerical experiments when replica symmetry is broken.
This is because the theoretical prediction calculates the expected
value of the fraction of frozen nodes in one (or a few) dominant
pure state, whereas, the landscape at the system sizes studied by
the EO algorithm has important finite size effects and strong sample
to sample fluctuations. We propose an argument that could explain
the convergence of the numerical results to the theoretical ones
from below: In many samples, additional minima that would be
negligible asymptotically are captured by EO, moreover, each minimum
is significantly distant from the other minima, and so unfreezes a
significant fraction of variables which are initially frozen in the
dominating minima (that we expect to agree with the theory). This
may also be one of reasons for the inconsistency of entropy values
between theory and EO statistics at low $\phi_{s}$ in
Fig.~\ref{entro}. The numerical results for the density of frozen
variables approximately agree with the RS prediction when the RS
entropy in Fig.~\ref{GS} remains positive, but deviate significantly
from theory in the strongly constrained regime, where the RS entropy
becomes negative.

The effects of frozen hubs on the breakdown of the RS prediction are
further illustrated in Fig.~\ref{substruct}. We classify the
deficient nodes in the network into $2$-core nodes and peripheral
nodes (those that are not on $2$-core). Figure~\ref{substruct} (a)
and (d) show that the fraction of $2$-core nodes deviates from the
RS prediction when $\phi_{s}$ is small, whereas the fraction of
frozen peripheral nodes is almost consistent with the RS prediction
in the entire range of $\phi_{s}$. Among the $2$-core nodes, we
further classify the deficient nodes into hubs (those connected to
$3$ other $2$-core nodes) and chain nodes (non-hubs connected to
only $2$ other $2$-core nodes) and compare their behaviors in
Fig.~\ref{substruct} (b) and (c). We conclude that the deviation
from the RS prediction is primarily attributed to the hubs rather
than other substructures as shown in Fig.~\ref{substruct}. There is
freezing close to the quenched boundary, which is due to local
RS-like features, and freezing on the core that relates to the
graph-wide RSB phenomena; the latter becomes increasingly important
as $\phi_s$ decreases. The strong deviation from theory indicates
that optimization problems with quenched disorder, as analyzed in
this paper, could provide an alternative test of theories of the
relationship between the freezing and the RSB
transition~\cite{ZH-prl05,WW-2012}.

\section{Numerical studies on single instances}
\label{sec_result}

In this section, we apply the entropic message passing approach
derived in Sec.~\ref{sec:RS} to study single instances of networks.
Two kinds of decimation strategy are used to identify the optimal
source location. One strategy is the maximal decimation in which the
most polarized nodes are fixed once the algorithm converges and then
the network is simplified correspondingly. This strategy can be
thought as a hard decimation since it is equivalent to adding an
external field of infinite intensity to the decimated
nodes~\cite{Brau-2005,Montanari-2007,Tersenghi-2009}. The other is
the reinforcement strategy which can be viewed as a sort of soft
decimation. In this strategy, the cavity probability is strengthened
or attenuated at each step by an external bias whose intensity is
updated with a rate increasing with the run
time~\cite{Chavas-2005,Zecchina-2008entropy,Lenka-08jstat}. When the
updating rate and the intensity of the external bias are correctly
chosen, the external messages are able to drive the iterations
towards some optimal assignment thanks to the fact that this soft
decimation utilizes the global information of the network at each
step. The belief propagation inspired decimation is also compared.
Hereafter, we use the short-hand notations BPD for belief
propagation inspired decimation, EMPD for entropic message passing
inspired decimation and EMPR for entropic message passing
reinforcement algorithm.

\subsection{Maximal decimation strategy}
For the maximal decimation strategy, we fix a fraction $f_{d}$ of
$N_{t}$ unfixed nodes with the largest full probability
$\psi_{i}^{s}$ to their most probable states once the algorithm
converges at the $t$-th sweep ($f_{d}=2\%$ in our simulations). Each
sweep consists of a sequential update of the messages on all the
edges of the network in a random order. When $f_{d}N_{t}<1$, only
one node with the largest bias is fixed. In the glassy phase, the
algorithm typically fails to converge, and this is due to the
building up of the long range correlations of different parts of the
network. Therefore, we carry out the decimation according to the
time-averaged full probability (over a number of sweeps) instead of
the instantaneous value of the probability~\cite{TABP-2006pre}. When
a node, say $i$ is fixed, the network is simplified by the following
decimation procedure. If the node is fixed to the source state, it
will send out the message $\psi_{i\rightarrow j}^{s}=1$ to all its
neighbors $j\in\partial i$ regardless of the later recursions.
Otherwise, all neighbors of node $i$ should take source state and
are fixed at the same time since no paired consumer nodes are
allowed in the singlet regime. Correspondingly, these fixed source
nodes send out a constant message $\psi_{i\rightarrow j}^{s}=1$ to
their neighbors. However, if at least one of neighbors of node $i$
has been fixed to consumer in this case, a contradiction will be
reported and we restart the algorithm. After the decimation, the
recursion is carried out on the unfixed nodes of the network until
$N_{t}=0$ and an optimal assignment of source location is obtained.
One can compute the marginal probability either based on the belief
propagation $\psi_{i\rightarrow
j}^{s}=\delta_{\Lambda_{i},A}+\delta_{\Lambda_{i},-1}\left[1-\prod_{k\in\partial
i\backslash j}\psi_{k\rightarrow i}^{s}\right]$ derived using the
energetic cavity method, or according to the entropic message
passing equations derived in Sec.~\ref{sec:RS}. The pseudocode of
EMPD is given as follows:

\begin{algo}
\small
\noindent\textbf{EMPD} algorithm \label{EMPD}\\
\noindent INPUT: the network with a fraction $\phi_{s}$ of surplus
nodes; a maximal number of iterations $T_{max}$; a predefined
precision $\eta$.\\
 \noindent OUTPUT: one assignment in the
singlet regime or 'probably no solutions'.
\begin{itemize}
\item[1.] Initialize randomly $\psi_{i\rightarrow j}^{s},\Delta S_{i\rightarrow j}$ for all edges of the network except for the edges
connected to the surplus nodes;
\item[2.] for $t=1$ to $t=T_{max}$ do:
\begin{itemize}
\item[2.1] a sweep of random sequential update of the messages
$\{\psi_{i\rightarrow j}^{s},\Delta S_{i\rightarrow j}\}$ for all
edges according to Eqs.~(\ref{cavfe04}),~(\ref{cavfe05})
and~(\ref{cavfe06});
\item[2.2]if $|\psi_{i\rightarrow j}^{s}(t)-\psi_{i\rightarrow j}^{s}(t-1)|<\eta$ and $|\Delta S_{i\rightarrow j}(t)-\Delta S_{i\rightarrow j}(t-1)|<\eta$
for all edges and the iteration has converged: goto line 3;
\end{itemize}
\item[3.] if $t<T_{max}$, compute the full probability $\psi_{i}^{s}$ for
unfixed nodes and decimate the network, else do the time average of
the full probability over the later $T_{max}/2$ sweeps and decimate
the network;
\item[4.] if an optimal assignment is found, return the solution and
stop; else if no contradiction is found, continue the decimation
procedure on the smaller network (goto line 2); else if a
contradiction is meet, return 'probably no solutions' and stop.
\end{itemize}
\end{algo}

\subsection{Soft decimation strategy}
The reinforcement strategy has been used to find solutions for
random constraint satisfaction
problems~\cite{Chavas-2005,Zecchina-2008entropy,Lenka-08jstat}.
Here, we introduce an external bias $\mu_{i}$ for each node, and its
intensity is updated with a probability increasing with the running
time as $1-t^{-r}$. The external bias $\mu_{i}$ is updated as
$\mu_{i}=\pi$ if $\psi_{i}^{s}<0.5$ and $1-\pi$ otherwise. The bias
strength $\pi\in[0,0.5]$. This implies that the current value of the
cavity probability of one node will be enhanced by a factor if its
full probability at the preceding iteration is biased towards the
source state. Thus the only modification to the original recursive
equation Eqs.~(\ref{cavfe04a}),~(\ref{cavfe05a})
and~(\ref{cavfe06a}) is
\begin{equation}\label{rmp}
     \tilde{\psi}_{i\rightarrow j}^{s}=\frac{\mu_{i}\psi_{i\rightarrow j}^{s}}{\mu_{i}\psi_{i\rightarrow j}^{s}+(1-\mu_{i})(1-\psi_{i\rightarrow j}^{s})}
\end{equation}
where $\tilde{\psi}_{i\rightarrow j}^{s}$ is the reinforced cavity
probability while $\psi_{i\rightarrow j}^{s}$ is the original one
computed from Eqs.~(\ref{cavfe04a}),~(\ref{cavfe05a})
and~(\ref{cavfe06a}). Both $\pi$ and $r$ should be optimized so as
to properly guide the iteration to converge to an optimal
assignment. In the numerical simulation, we fix $r=0.1$ although
other choices also lead to find a solution, e.g., $r=0.3$. The
algorithm is precisely described as follows:

\begin{algo}
\small
\noindent\textbf{EMPR} algorithm \label{EMPR}\\
\noindent INPUT: the network with a fraction $\phi_{s}$ of surplus
nodes; a maximal number of iterations $T_{max}$; two empirical parameters $\pi$ and $r$.\\
 \noindent OUTPUT: one assignment in the
singlet regime or 'probably no solutions'.
\begin{itemize}
\item[1.] Initialize randomly $\psi_{i\rightarrow j}^{s},\Delta S_{i\rightarrow j}$ for all edges of the network except for the edges
connected to the surplus nodes and initialize the external biases
$\{\mu_{i}\}$ at random;
\item[2.] for $t=1$ to $t=T_{max}$ do:
\begin{itemize}
\item[2.1] a sweep of random sequential update of the messages
$\{\psi_{i\rightarrow j}^{s},\Delta S_{i\rightarrow j}\}$ for all
edges according to
Eqs.~(\ref{cavfe04}),~(\ref{cavfe05}),~(\ref{cavfe06}) and
Eq.~(\ref{rmp});
\item[2.2]update all the external biases with probability $1-t^{-r}$;
\item[2.3]assign $s_{i}=1$ if $\mu_{i}>0.5$ and $0$ otherwise; if
$\{s_{i}\}$ is an optimal assignment (no contradiction), return the
solution and stop.
\end{itemize}
\item[3.] if $t=T_{max}$, return 'no solutions at current values of $\pi$ and
$r$'.
\end{itemize}
\end{algo}

\begin{figure}
\centering
    \includegraphics[bb=14 15 287 217,scale=1.0]{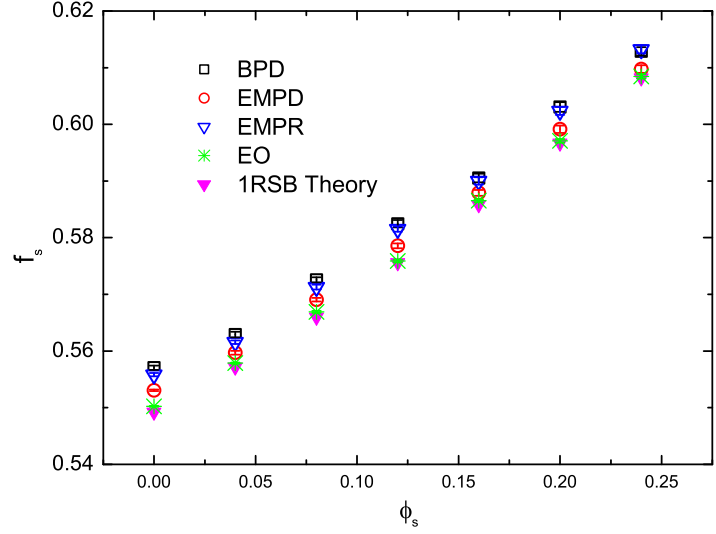}
  \caption{
  (Color online) The fraction of source nodes in the optimized
  network ($N=2000$) obtained by different algorithms. For BPD and EMPD, $f_{d}=2\%$, $T_{max}=2000$ and $\eta=10^{-3}$ and additionally for EMPD, $\varepsilon=10^{-6}$.
  $T_{max}=10^{4}$ for EMPR. $\tau=1.8$ for EO. The data point is the mean over $30$ random regular networks with $C=3$ and the error bar is also shown.
  }\label{Alg}
\end{figure}

The inference results using the above proposed algorithms on single
instances of size $N=2000$ are compared in Fig.~\ref{Alg}. BPD has
been compared with GSAT algorithm which randomly selects a small
cluster of nodes and then update its configuration to the one with
lowest network energy, however, GSAT yields a higher
energy~\cite{Wong-2010}. As shown apparently in Fig.~\ref{Alg}, the
entropic message passing algorithm yields a lower value of $f_{s}$
than the belief propagation without taking the entropic effects into
account, and hence achieves a more optimal cost. Its value is also
closer to the theoretical $1$RSB prediction. This indicates the
advantages of employing entropic information in message passing
procedure to differentiate states that are locally ambiguous but
possibly lead to frustrations in the long range.

It should be mentioned that EMPR can find the optimal assignments
faster than other strategies in the glassy regime. This is because,
on the one hand, the presence of an updating external bias manages
to drive the evolution of the reinforced cavity probabilities
towards the ground state where the satisfying assignment is
maximally aligned with the externally imposed direction; on the
other hand, the hard decimation fails to converge in this regime and
the time-average of the marginal probabilities is required, which
increases the time complexity of the algorithm. In the RS region,
EMPR becomes slow because the proper values of $(\pi,r)$ could not
be easily found, probably due to the presence of the many background
messages sent out by the surplus nodes. In contrast, the EMPD and
BPD are typically convergent in this regime and thus fast to
identify the optimal assignment. In the glassy regime, the estimated
$f_{s}$ by message passing algorithms is higher than the ground
state value predicted by the $1$RSB theory. The physical
interpretation is, in the glassy region, exponentially many
metastable states act as dynamical arrests for various simple
heuristics. Furthermore, the ground state predicted by the $1$RSB
theory can be unstable towards further steps of RSB or full RSB (an
infinite hierarchy of nested states)~\cite{Wong-2010,MM-2009}, and
the lower bound to the true dynamical threshold of $f_{s}$ is
predicted by the Gardner value~\cite{Gardner-1985} above which the
$1$RSB metastable states become unstable~\cite{Instability-2004}.
However, the entropic message passing algorithm on single instances
yields a lower value of $f_{s}$, and particularly the EMPD can give
a lowest value of $f_{s}$ among all message passing algorithms in
the glassy region. Note that the attained energy can be further
lowered for the hard decimation by taking smaller $f_{d}$ at the
expense of larger computational time. We also put the inference
result by EO algorithm (see Appendix~\ref{sec:EO}) in
Fig.~\ref{Alg}. The time complexity for EO to find a ground state is
of the order $\mathcal{O}(N^{3})$ while the hard decimation
converges in $\mathcal{O}(N^{2})$ when fixing at each step only one
variable and the soft decimation converges in the order of
$\mathcal{O}(N)$. When we fix a given fraction of variables at each
step, the time complexity for the hard decimation reduces to be of
the order $\mathcal{O}(N\ln N)$.

\section{Conclusion}
\label{sec_Sum}

We have considered the entropic effects in the ground state of the
source location problem and derived the associated entropic message
passing algorithms as an improvement over the previous energetic
computation of the problem. Using the $1$RSB ansatz, the ground
state entropy predicted by the RS solution is improved and yields a
better approximation. Although the formula for the entropy in the RS
ansatz is equivalent to that of the minimal vertex cover problem,
computation of the entropy in the $1$RSB ansatz requires additional
energetic information different from that in the vertex cover
problem. In particular, the energy changes we used to compute the
entropy values in the $1$RSB case include the influence of the
forward links, which is necessary for the case where flows on the
links also contribute to the energy. We found that the $1$RSB
recursions can be computed in a way similar to that in
Ref.~\cite{Wong-2010} which did not consider the entropy but only
the energy, but by carefully monitoring the free energy changes due
to restoration and reconnection of nodes and links, we found an
alternative derivation. This also provides tools to analyze network
optimization problems dependent on both node and link conditions.

The predicted $f_{s}$ are checked by the simulations on single
instances of transportation networks. Using the message passing
inspired decimation algorithms, we find that the entropic
information helps to lower the inference value of $f_{s}$ making it
closer to the ground state value. This advantage is due to the extra
information gained from entropy considerations guiding us to choose
one of the two possible bistable states correctly and hence resolve
potential conflicts arising from long-range frustrations.

The theoretical results of ground state energy agree with those
obtained independently by the EO method. However, the agreement of
the entropy value at low $\phi_{s}$ is less satisfactory. Large
finite size effects and strong sample-to-sample fluctuations are
present. We also found that the entropy is extensive when the RS
prediction becomes unstable, matching theoretical predictions in the
$1$RSB framework. By studying the fraction of frozen nodes according
to their topological classification, it was revealed that frozen
hubs are the primary cause for the breakdown of the RS prediction in
its unstable regime. On the technical aspect, large size
measurements of frozen set of variables and entropies are useful in
understanding the RSB phenomena, but the sampling of ground states
is numerically challenging.

As the installation cost parameter $u$ increases, another important
configuration of source and consumer nodes, namely the doublet
regime will appear. Extension of the current method to this regime
would be very interesting. In this case, to derive the recursive
equations, we need to define two extra cavity messages, one for
singly consuming state and the other for doubly consuming state. Due
to many more ways of clustering the consumer nodes, we anticipate
that entropic effects will be very significant at the commensurate
points in the doublet regime~\cite{Wong-2010}. The present work has
established the foundation, and one can extend the analysis by
including one more term for the doubly consumer state in the free
energy expression. It is also of interest to extend the current
analysis to the lattice glass models~\cite{Biroli-02prl,Mezard-04}
where for example each site in the (finite dimensional or Bethe)
lattice can have at most one particle and any particle has at most a
fixed number of occupied nearest neighbor. Other possible
applications may be found in routing and path selection problems on
sparse graphs~\cite{CH-12}.

\section*{Acknowledgments}
 We thank Dr. Chi Ho Yeung for helpful discussions on the alternative derivations of the $1$RSB entropy. This work was partially supported by the Research Council of Hong Kong (Grant Nos. HKUST 605010 and 604512).
\appendix
\section{Cavity analysis of the model}
\label{sec:CAV} Here we briefly present a theoretical analysis of
the model defined in Eq.~(\ref{cost2}) based on the cavity
method~\cite{MM-2009} and finally demonstrate its relation to the
minimal vertex cover problem. The original derivation was given in
Ref.~\cite{Wong-2010}.

The network we consider here has a locally tree-like structure, as
described in Fig.~\ref{ebpfig2} or Fig.~\ref{site}. We define
$E_{i\rightarrow j}(x_{ji})$ as the cavity energy of the tree
terminated at node $i$ without consideration of its ancestor node
$j$, and $E_{i\rightarrow j}(x_{ji})$ is determined by
\begin{equation}\label{cavE}
\begin{split}
E_{i\rightarrow j}(x_{ji})&=\mathcal {E}(E_{k_1\rightarrow
i},E_{k_2\rightarrow i},\ldots,E_{k_{C-1}\rightarrow
i};\Lambda_i,x_{ji})\\
&\equiv\min_{\{x_{ik}\}}\left[\sum_{k\in\partial i\backslash
j}E_{k\rightarrow
i}(x_{ik})+\frac{u^{2}}{2}\Theta\left(-\Lambda_i-\sum_{k\in\partial
i\backslash j}x_{ik}+x_{ji}\right)+\frac{x_{ji}^{2}}{2}\right],
\end{split}
\end{equation}
in the zero-temperature limit. The first term sums over all energies
of descendants, while the second and last terms refer to the penalty
for negative final resource and transportation cost respectively. To
write a recursion of an intensive energy, we separate the extensive
quantity $E_{i\rightarrow j}(x_{ji})$ into two terms:
\begin{equation}\label{intE}
    E_{i\rightarrow j}(x_{ji})=E_{i\rightarrow j}^{V}(x_{ji})+E_{i\rightarrow
    j}(0),
\end{equation}
where $E_{i\rightarrow j}^{V}(x_{ji})$ is called a vertex-dependent
intensive energy such that $E_{i\rightarrow j}^{V}(0)=0$. In fact,
$E_{i\rightarrow j}^{V}(x_{ji})$ describes the energy variation from
$E_{i\rightarrow j}(0)$ as $x_{ji}$ changes. This allows us to
recast Eq.~(\ref{cavE}) into
\begin{equation}\label{recE}
E_{i\rightarrow j}^{V}(x_{ji})=\mathcal {E}(E_{k_1\rightarrow
i}^{V},E_{k_2\rightarrow i}^{V},\ldots,E_{k_{C-1}\rightarrow
i}^{V};\Lambda_i,x_{ji})-\mathcal {E}(E_{k_1\rightarrow
i}^{V},E_{k_2\rightarrow i}^{V},\ldots,E_{k_{C-1}\rightarrow
i}^{V};\Lambda_i,0).
\end{equation}
From the above definition, one arrives at the energy change due to
addition of a node:
\begin{equation}\label{Esite}
    \Delta E_{{\rm node}}=\mathcal {E}(E_{k_1\rightarrow
i}^{V},E_{k_2\rightarrow i}^{V},\ldots,E_{k_{C}\rightarrow
i}^{V};\Lambda_{{\rm node}},0),
\end{equation}
and the energy change due to addition of a link between nodes $L$
and $R$:
\begin{equation}\label{Elink}
    \Delta
    E_{{\rm link}}=\min_{x}\left[E_{L}^{V}(x)+E_{R}^{V}(-x)-\frac{x^{2}}{2}\right].
\end{equation}
Finally, the typical energy per node (energy density) is given by
$\left<E\right>=\left<\Delta E_{{\rm
node}}\right>-\frac{C}{2}\left<\Delta E_{{\rm
link}}\right>$~\cite{cavity-2001}.

Solving Eq.~(\ref{recE}) is in general infeasible. However, the form
of Eq.~(\ref{cost2}) implies a piecewise quadratic expression for
$E^{V}$, which greatly simplifies our analysis. We can write the
cavity energy functions as composite functions
\begin{equation}\label{compFunc}
   E_{i\rightarrow j}^{V}(x_{ji})=\min_{n_{i\rightarrow j}}[f_{n_{i\rightarrow
   j}}(x_{ji})],
\end{equation}
where $f_{n_{i\rightarrow
   j}}(x)$ is a quadratic function of the form
\begin{equation}\label{compFunc02}
   f_n(x)=a_n(x-\tilde{x}_n)^{2}+d_n.
\end{equation}
The state $n=0$ represents the {\em s}-state and $n\geq1$ represents
various consumer states. Thus, for $n=0$, $a_0=1/2,\tilde{x}_0=0$
and
\begin{equation}\label{DFunc}
  d_{n_{i\rightarrow j}=0}=\frac{u^{2}}{2}+\sum_{k\in\partial i\backslash
  j}d_{n^{*}_{k\rightarrow i}}-\Delta E_{i\rightarrow j},
\end{equation}
where $\{n^{*}_{k\rightarrow i}\}$ is the set of $n_{k\rightarrow
i}$ that minimizes $d_{n_{i\rightarrow j}=0}$ and $\Delta
E_{i\rightarrow j}\equiv E_{i\rightarrow j}(0)-\sum_{k\in\partial
i\backslash j}E_{k\rightarrow i}(0)$ is the cavity energy change.
For node $i$ being the {\em c}-state, the cavity equations read
\begin{subequations}\label{compEq}
\begin{align}
  a_{i\rightarrow j}&=\frac{1}{2}\left[1+\frac{1}{\sum_{k\in\partial i\backslash j}(2a_{k\rightarrow i})^{-1}}\right],\\
  \tilde{x}_{i\rightarrow j}&=\frac{-1+\sum_{k\in\partial i\backslash j}\tilde{x}_{k\rightarrow i}}{1+\sum_{k\in\partial i\backslash j}(2a_{k\rightarrow i})^{-1}},\\
  d_{n_{i\rightarrow j}}&=\frac{\left(-1+\sum_{k\in\partial i\backslash j}\tilde{x}_{k\rightarrow i}\right)^{2}}{2\left[1+\sum_{k\in\partial i\backslash j}(2a_{k\rightarrow i})^{-1}\right]}
  +\sum_{k\in\partial i\backslash j}d_{n_{k\rightarrow i}}-\Delta E_{i\rightarrow j}.
\end{align}
\end{subequations}

In the singlet regime for networks with fixed connectivity $C$, we
only need to consider the case that all nodes $k\in\partial
i\backslash j$ are in the {\em s}-state. In this case,
$a_{i\rightarrow j}=C/(2C-2),\tilde{x}_{i\rightarrow j}=-1/C$, and
$d_{n_{i\rightarrow j}}=1/(2C)+\sum_{k\in\partial i\backslash
j}d_{n_{k\rightarrow i}=0}-\Delta E_{i\rightarrow j}$. Other
combinations of the states of $k\in\partial i\backslash j$ yield
higher energies and can be ignored. Since the coefficients
$a_{i\rightarrow j}$ and $\tilde{x}_{i\rightarrow j}$ are fixed, the
recursion relations can be further simplified to those of the energy
minima $d_{n_{i\rightarrow j}=0}$ and $d_{n_{i\rightarrow j}=1}$,
where
\begin{subequations}\label{DFunc02}
\begin{align}
   d_{n_{i\rightarrow j}=0}&=\frac{u^{2}}{2}
  +\sum_{k\in\partial i\backslash j}\min(d_{n_{k\rightarrow i}=0},d_{n_{k\rightarrow i}=1})-\Delta E_{i\rightarrow
  j},\\
 d_{n_{i\rightarrow j}=1}&=\frac{1}{2C}
  +\sum_{k\in\partial i\backslash j}d_{n_{k\rightarrow i}=0}-\Delta E_{i\rightarrow
  j}.
  \end{align}
\end{subequations}
To determine the cavity states, it is sufficient to consider the
energy difference $\epsilon_{i\rightarrow j}\equiv
d_{n_{i\rightarrow j}=1}-d_{n_{i\rightarrow j}=0}$, given by
Eq.~(\ref{recSimp}) in the main text.

Furthermore, derivation in Sec.~\ref{RS01} gives back not only the
ground state entropy but also the ground state energy (see also
Fig.~\ref{ebpfig2}). Due to this feature of the model, one should
consider the flow on the forward link when determining the cavity
state of a node, which leads to the analysis in Sec.~\ref{RS03} and
has significant implications for further analysis of more complex
connection patterns when $u$ increases to a higher value. Note also
that the cavity energy derived in Sec.~\ref{RS03} can be used to
calculate the reweighting factor in $1$RSB equations~(\ref{SPy}),
which is different from the case in the minimal vertex cover
problem~\cite{ZH-pre06}, particularly when metastable states with
high-lying energies are considered during the cavity iterations. In
the singlet regime, Eq.(\ref{cost2}) describes the same set of
ground state configurations as Eq.(\ref{EO-1}), but the $1$RSB
picture of the ground state entropy in the thermodynamic limit can
only be obtained using both cavity energy and entropy information
presented in the main text. More explanations are also given in
Appendix~\ref{sec:EO}.

\section{Alternative derivation of entropic message passing equations}
\label{sec:appendix} In this appendix, we present an alternative
derivation of the entropic message passing equations in
Sec.~\ref{sec:RS}. The recursive relations for the cavity
probability and entropy are obtained in a probabilistic way by
focusing on the change of the ground state size under the cavity
iterations~\cite{ZH-pre09}. The value for the cavity probability
$\psi_{i\rightarrow j}^{s}$ of node $i$ depends on the incoming
cavity probabilities from its neighbors other than node $j$, which
can be categorized into three cases~\cite{Wong-2010}.

In the first case as depicted in Fig.~\ref{site} (a), all neighbors
of node $i$ have non-zero cavity probabilities $\{\psi_{k\rightarrow
i}^{s}\}$. In this case, the cavity state of node $i$ must be a
consumer. We assume that the number of optimal assignments before
addition of node $i$ is $\Omega_{N-1}$. After the node addition in
Fig.~\ref{site} (a), $\Omega_{N-1}$ should be reduced since node $i$
is now in the consumer state and should remain as a singlet. Hence
all its neighbors other than $j$ should take source state and only
those configurations with $s_{k}=1$ $(k\in\partial i\backslash j)$
in the ground state of the system with $N-1$ nodes are valid after
the addition of node $i$. In this case, $\psi_{i\rightarrow
j}^{s}=0$, and the number of optimal assignments
$\Omega_{N}=\Omega_{N-1}\prod_{k\in\partial i\backslash
j}\psi_{k\rightarrow i}^{s}$ where the product comes from the weak
correlation assumption. The cavity entropy change is readily
obtained as
\begin{equation}\label{Ecm01}
        \Delta S_{i\rightarrow
        j}=\ln\frac{\Omega_{N}}{\Omega_{N-1}}=\sum_{k\in\partial i\backslash
        j}\ln\psi_{k\rightarrow i}^{s}.
\end{equation}

The second case (Fig.~\ref{site} (b)) gets a bit more involved. In
this case, only one neighbor of node $i$, say node $k$, is frozen to
the consumer state in the absence of node $i$, i.e.,
$\psi_{k\rightarrow i}^{s}=0$. According to Fig.~\ref{site} (b), the
outcome is that the cavity source and consumer states of node $i$
are degenerate. In Ref.~\cite{Wong-2010}, node $i$ in this case was
treated as the so-called bistable node. To consider the entropy
change, we note that before the addition of node $i$, the number of
optimal assignments in the ground state is
$\Omega_{N-1}=\Omega_{N-2}\prod_{k'\in\partial k\backslash
i}\psi_{k'\rightarrow k}^{s}$ where $\Omega_{N-2}$ is the number of
optimal assignments in the ground state of the network without node
$k$ and $i$. Note that when node $k$ is frozen to the consumer
state, the same constraint as the first case, namely, that
$\psi_{k'\rightarrow k}^{s}>0$ for all $k'\in\partial k\backslash
i$, is already imposed on node $k$. Following the discussion on the
first case, this also implies that $\Omega_{N-1}$ can be simplified
to $\Omega_{N-1}=\Omega_{N-2}e^{\Delta S_{k\rightarrow i}}$ where
$\Delta S_{k\rightarrow i}=\sum_{k'\in\partial k\backslash
i}\ln\psi_{k'\rightarrow k}^{s}$. The following two paragraphs
analyze separately the possibilities that node $i$ takes the source
or consumer states.

If node $i$ takes the source state as shown in the middle panel of
Fig.~\ref{site} (b), node $k$ needs not change its state, therefore
the number of optimal assignments after node addition with $s_{i}=1$
is the same as $\Omega_{N-1}$. Hence we have
$\Omega_{N}|_{s_{i}=1}=\Omega_{N-2}e^{\Delta S_{k\rightarrow i}}$.

However, if node $i$ chooses the consumer state, then the node $k$
should change its state from consumer to source since we focus on
the singlet regime where no paired consumer nodes are allowed. This
change does not impose any further restrictions on the set of its
neighbors $\partial k\backslash i$, since the neighbors of a source
node can either be sources or consumers. On the other hand,
assigning node $i$ to be in the consumer state restricts all its
neighbors other than $k$ to be in source states only. Consequently,
the number of optimal assignments after node addition with $s_{i}=0$
is $\Omega_{N}|_{s_{i}=0}=\Omega_{N-2}\prod_{l\in\partial
i\backslash k,j}\psi_{l\rightarrow i}^{s}$ where node $i$ is fixed
to the consumer state (see the right panel of Fig.~\ref{site} (b)).

Since node $i$ has the choices to be in those ground state
assignments that have $s_{i}=0$ or $s_{i}=1$, the total number of
optimal assignments after addition of node $i$ is
$\Omega_{N}=\Omega_{N-2}e^{\Delta S_{k\rightarrow
i}}+\Omega_{N-2}\prod_{l\in\partial i\backslash
k,j}\psi_{l\rightarrow i}^{s}$. The associated entropy change can be
expressed as
\begin{equation}\label{Ecm02a}
        \Delta S_{i\rightarrow
        j}=\ln\frac{\Omega_{N}}{\Omega_{N-1}}=\ln\left[1+e^{-\Delta S_{k\rightarrow i}}\prod_{l\in\partial i\backslash
k,j}\psi_{l\rightarrow i}^{s}\right].
\end{equation}
At the same time, the cavity probability $\psi_{i\rightarrow j}^{s}$
is determined by
\begin{equation}\label{Ecm02b}
\begin{split}
        \psi_{i\rightarrow
        j}^{s}&=\frac{\Omega_{N}|_{s_{i}=1}}{\Omega_{N}|_{s_{i}=1}+\Omega_{N}|_{s_{i}=0}}\\
&=\frac{1}{1+e^{-\Delta S_{k\rightarrow i}}\prod_{l\in\partial
i\backslash k,j}\psi_{l\rightarrow i}^{s}}.
\end{split}
\end{equation}

The third case where at least two of incoming $\psi_{k\rightarrow
i}^{s}$ for node $i$ vanish is presented in Fig.~\ref{site} (c). The
added node $i$ should take the source state, i.e.,
$\psi_{i\rightarrow j}^{s}=1$. In Fig.~\ref{site} (c), both
$\psi_{k\rightarrow i}^{s}$ and $\psi_{l\rightarrow i}^{s}$ are
equal to zero, thus the number of optimal assignments before the
node addition is $\Omega_{N-1}=\Omega_{N-3}\prod_{k'\in\partial
k\backslash i}\psi_{k'\rightarrow k}^{s}\prod_{l'\in\partial
l\backslash i}\psi_{l'\rightarrow l}^{s}$ where $\Omega_{N-3}$ is
the number of optimal assignments in the ground state of the network
without node $k$,$l$ and $i$. After node $i$ is added, node $i$ is
frozen to the source state and the neighbors $k$ and $l$ need not
change their states, as a result, the number of optimal assignments
$\Omega_{N}$ after addition of node $i$ is identical to
$\Omega_{N-1}$. We conclude that the cavity entropy change for the
third case is zero.

The full entropy change on adding a node $i$ to the network can be
derived by extending the above analysis to cover all neighbors of
the node. Hence the expression of $\Delta S_{i}$ is given by
Eqs.~(\ref{Ecm01}) and~(\ref{Ecm02a}) in the first and second cases
respectively, except that the neighboring set $\partial i\backslash
j$ is replaced by $\partial i$, and $\Delta S_{i}=0$ in the third
case.

 To obtain the entropy density of the network, we need to add the
 link between two randomly selected nodes and consider the entropy
 change due to this link addition. This includes two cases as
 depicted in Fig.~\ref{edge} (a) and (b) respectively. In the
 first case where at most one end of the link takes positive cavity
 probability, then after the link addition, those configurations
 where both node $i$ and $j$ take the consumer state should be
 excluded from the ground state whose size is denoted by $\Omega$, therefore, the
number of the optimal
 assignments in the current ground state should be $\Omega'=\Omega-\Omega(1-\psi_{i\rightarrow j}^{s})(1-\psi_{j\rightarrow
 i}^{s})$ with the entropy change $\Delta S_{\left(ij\right)}=\ln\left[1-(1-\psi_{i\rightarrow j}^{s})(1-\psi_{j\rightarrow
 i}^{s})\right]$. In the second case where both ends of the added
 link are frozen into the consumer state before the link addition, the
 number of optimal assignments in the ground state without the link
 is $\Omega=\Omega_{N-2}\prod_{l\in\partial i\backslash j}\psi_{l\rightarrow i}^{s}\prod_{l'\in\partial j\backslash i}\psi_{l'\rightarrow
 j}^{s}=\Omega_{N-2}e^{\Delta S_{i\rightarrow j}}e^{\Delta S_{j\rightarrow i}}$ where $\Omega_{N-2}$ is the number of optimal assignments
 in the ground state without node $i$ and $j$. After the link
 addition, either node $i$ or node $j$ changes its state to the source
 state. The number of optimal assignments in the current ground
 state becomes $\Omega'=\Omega'|_{s_{i}=1}+\Omega'|_{s_{j}=1}$ where $\Omega'|_{s_{i}=1}=\Omega_{N-2}\prod_{l'\in\partial j\backslash i}\psi_{l'\rightarrow
 j}^{s}=\Omega_{N-2}e^{\Delta S_{j\rightarrow i}}$ and $\Omega'|_{s_{j}=1}=\Omega_{N-2}\prod_{l'\in\partial i\backslash j}\psi_{l'\rightarrow
 i}^{s}=\Omega_{N-2}e^{\Delta S_{i\rightarrow j}}$. Thus the entropy change due to the link addition in the
 second case is $\Delta S_{\left(ij\right)}=\ln\left[e^{-\Delta S_{i\rightarrow j}}+e^{-\Delta S_{j\rightarrow
 i}}\right]$. To sum up, the entropy change due to the
 edge addition is written as
\begin{equation}\label{Ecm04app}
\begin{split}
        \Delta S_{\left(ij\right)}&=\Theta(\psi_{i\rightarrow j}^{s}+\psi_{j\rightarrow
        i}^{s})\ln\left[1-(1-\psi_{i\rightarrow j}^{s})(1-\psi_{j\rightarrow
 i}^{s})\right]\\
 &+\left[1-\Theta(\psi_{i\rightarrow j}^{s}+\psi_{j\rightarrow
        i}^{s})\right]\ln\left[e^{-\Delta S_{i\rightarrow j}}+e^{-\Delta S_{j\rightarrow
 i}}\right].
 \end{split}
\end{equation}

\section{Removing and restoring a node and a link} \label{app:RS02}

For a network with $N$ nodes and $L$ links, we consider an initial
configuration with $N-1$ nodes and $L-C$ links, obtained by removing
node $i$ and its adjacent links. Note that for each node
$k\in\partial i$, the forward link $k\rightarrow i$ is no longer
present, so that the flow $x_{ik}$ is no longer considered in
optimizing the cavity energy of node $k$. Now we consider the energy
of node $k$ taking the {\em c} state when all neighboring nodes in
the set $\partial k\backslash i$ take the {\em s} state. Each of
these neighbors provides a flow of $1/(C-1)$ to node $k$, so that
the transportation cost becomes $1/(2C-2)$. In the singlet regime,
this is higher than the energy of node $k$ taking the {\em s} state,
which is $u^{2}/2$. Hence in the low temperature limit, the initial
free energy only consists of contributions from the nodes $k$ taking
the {\em s} states, given by
\begin{equation}\label{rest01}
    F_{N-1,L-C}=-T\sum_{k\in\partial i}\ln Z_{k\rightarrow i}^{s}.
\end{equation}
Then we consider the final free energy after the node $i$ and all
its adjacent links are restored to this configuration as shown in
Fig.~\ref{resrec} (e). Extending Eq.~(\ref{cavfe01}) to include all
neighbors of node $i$, we have
\begin{equation}\label{rest02}
    F_{N,L}=\frac{u^{2}}{2}-T\ln\left[\prod_{k\in\partial i}(Z_{k\rightarrow i}^{s}+Z_{k\rightarrow i}^{c})+e^{\beta\gamma}
        \prod_{k\in\partial i}Z_{k\rightarrow i}^{s}\right].
\end{equation}
Hence the free energy change on restoring node $i$ and its links is
given by
\begin{equation}\label{rest03}
    \Delta F_{i}^{{\rm res}}=\frac{u^{2}}{2}-T\ln\left[1+e^{\beta\gamma}
        \prod_{k\in\partial i}\psi_{k\rightarrow
        i}^{s}\right]+T\sum_{k\in\partial i}\ln\psi_{k\rightarrow i}^{s}.
\end{equation}
Eq.~(\ref{rest03}) is derived by subtracting Eq.~(\ref{rest01}) from
Eq.~(\ref{rest02}) and using the definition of Eq.~(\ref{cavfe03}).
The entropy change $\Delta S_{i}^{\rm res}$ of restoring node $i$
and its links can then be computed in the zero temperature limit as
\begin{equation}\label{rest04}
\Delta S_{i}^{{\rm res}}=\Delta S_{i}^{{\rm rec}}+\sum_{l\in\partial
i}\Delta S_{l\rightarrow i}.
      \end{equation}
To obtain the last term of Eq.~(\ref{rest04}), we have used
$\psi_{l\rightarrow i}^{s}=e^{-\Delta S_{l\rightarrow i}}$ and
Eq.~(\ref{cavfe04a}) for non-vanishing and vanishing input cavity
probabilities respectively.

The cavity free energy change akin to a restoration process can  be
defined according to the node-headed diagrams in Fig.~\ref{resrec}
(d). In this case, the flow energy in the forward link $i\rightarrow
j$ is excluded, and the cavity free energy change in this {\em
restoration} case can be expressed as
\begin{equation}\label{restFe}
\Delta F_{i\rightarrow j}^{{\rm
res}}=\frac{u^{2}}{2}+T\sum_{k\in\partial i}\ln\psi_{k\rightarrow
i}^{s},
\end{equation}
 from which the ground state cavity
energy change computed in Ref.~\cite{Wong-2010} can be recovered.

To obtain the entropy contribution of an edge, we consider an
initial configuration with $N$ nodes and $L-1$ links, obtained by
removing the link between nodes $i$ and $j$. The initial free energy
is given by
\begin{equation}\label{rest05}
     F_{N,L-1}=u^{2}-T\ln\left[\prod_{k\in\partial i\backslash
j}(Z_{k\rightarrow i}^{s}+Z_{k\rightarrow
i}^{c})\right]-T\ln\left[\prod_{l\in\partial j\backslash
i}(Z_{l\rightarrow j}^{s}+Z_{l\rightarrow j}^{c})\right].
\end{equation}
Now we consider the final free energy after the link between nodes
$i$ and $j$ is added back to this configuration as shown in
Fig.~\ref{resrec} (f). Following the analysis of the reconnection
process in Sec.~\ref{RS03}, we analyze the free energy change
starting from the network with $N-2$ nodes obtained by excluding
nodes $i$ and $j$ and the link between them. This leads to the
following free energy change
\begin{equation}\label{rest06}
      \Delta F_{(ij)}^{{\rm res}}=\Delta F_{(ij)}^{{\rm rec}}-T\ln\left[1+e^{\beta\gamma}
        \prod_{k\in\partial i\backslash j}\psi_{k\rightarrow
        i}^{s}\right]-T\ln\left[1+e^{\beta\gamma}
        \prod_{l\in\partial j\backslash i}\psi_{l\rightarrow
        j}^{s}\right].
\end{equation}
We have used Eq.~(\ref{rest05}) and the definition of the cavity
probability $\psi_{i\rightarrow j}^{s}$ in Eq.~(\ref{cavfe03}) to
derive Eq.~(\ref{rest06}). Taking the zero temperature limit, we
obtain
\begin{equation}\label{rest07}
        \Delta S_{\left(ij\right)}^{{\rm res}}=\Delta S_{\left(ij\right)}^{{\rm rec}}+\Delta S_{i\rightarrow j}+\Delta S_{j\rightarrow i}.
\end{equation}

Comparing Eq. (\ref{rest04}) with the entropy change in the
reconnection process and Appendix~\ref{sec:appendix}, we see that
extra cavity entropy terms are present in the expression for
restoration. Similarly, comparing Eq. (\ref{rest07}) with the
entropy change in the reconnection process and
Appendix~\ref{sec:appendix}, we find additional cavity entropy
terms. This is due to the fact that the contributions of the forward
links are excluded before restoration, while they are included
before reconnection, as evident from a comparison between
Figs.~\ref{resrec} (b) and (e), and between Figs.~\ref{resrec} (c)
and (f). This difference is a consequence of the distribution of
energy among both nodes and links in the source location problem.

\section{EO algorithm to analyze ground state energy and entropy}
\label{sec:EO} In this section, we briefly introduce a stochastic
local search algorithm named EO algorithm~\cite{Boet-2003,Boet-2004}
to study the statistics of ground state energy and entropy for
moderate network size. For convenience in the following analysis, we
take $s_{i}=-1$ for source nodes and $+1$ for consumer nodes. In the
singlet regime, we write an energy cost for the algorithm to
minimize as
\begin{equation}\label{EO-1}
    \mathcal
    {H}(\mathbf{s})=\sum_{i}\frac{1-s_{i}}{2}+\sum_{(ij)}\frac{(1+s_{i})(1+s_{j})}{2}.
\end{equation}
The first term penalizes configurations with excess resource
suppliers and the second term penalizes the link whose both ends are
occupied by resource consumers. Thus minimizing $\mathcal {H}$ is
equivalent to finding the singlet state with maximal number of
consumers. As explained in Sec.~\ref{sec_SLP}, it is always strictly
energetically favorable in the singlet regime for two consumers to
form a consumer-supplier pair, thus Eq.~(\ref{EO-1}) describes the
same set of ground states of Eq.~(\ref{cost2}) in the {\em singlet}
regime, but disagrees on all the excitation levels. This new
Hamiltonian has discrete energy levels, and high degeneracy of local
energy levels, which makes the ranking subroutine in EO more
efficient. The marginal probability of $s_{i}$, given all other
variables, is described by
\begin{equation}\label{EO-2}
    P(s_{i}|\mathbf{s}\backslash s_{i})\propto \exp\left[\sum_{s}\beta
    h_{i}^{s}\delta_{s_{i},s}\right],
\end{equation}
where the local field in the ground state has only energetic
content, and is given by $h_{i}^{-1}=-1$ and
$h_{i}^{1}=-\sum_{j\in\partial i}(1+s_{j})$. Thus we can define
fitness used in the EO algorithm as
\begin{equation}\label{EO-3}
   f_{i}=\frac{s_{i}}{2}\sum_{s}sh_{i}^{s}.
\end{equation}
Thus a node with high fitness is favored to minimize the energy cost
in Eq.~(\ref{EO-1}). To find the ground state configurations, the EO
procedure ranks fitness for all nodes and then determines which spin
$s_{i}$ should be flipped, which is described by the following
implementation~\cite{Raymond-2012}.

\begin{algo}
\small
\noindent\textbf{EO} algorithm \label{EO}\\
\noindent INPUT: the network with a fraction $\phi_{s}$ of surplus
nodes; a maximal number of iterations $T_{E}$; the power law exponent $\tau$.\\
 \noindent OUTPUT: ground state energy.
\begin{itemize}
\item[1.] Initialize randomly $s_{i}$ for all node $i$ and set the corresponding energy $E_{0}$;
\item[2.] for $t=1$ to $t=T_{E}$ do:
\begin{itemize}
\item[2.1] Select at random a variable according to its fitness rank (probability of selection follows a power law like $k^{-\tau}$ where $k$ is the variable's rank);
\item[2.2] Flip the selected variable and update its fitness, and the fitness of its neighbors;
\item[2.3] Rerank the flipped variable and its neighbors;
\item[2.4] Update the energy; if the new energy is lower than $E_{0}$, reset $E_{0}$ to be the new energy.
\end{itemize}
\end{itemize}
\end{algo}
After $T_{E}$ updates, we have searched a fraction of the
configuration space, with a strong bias towards low energy states.
If $\tau$ is well chosen and $T_{E}$ is large enough, we avoid being
trapped by a local minimum, but manage to explore many such minima.

$\tau$ should be fixed in the range $1<\tau<2$, closer to one for
more challenging energy landscapes, based on theoretical
arguments~\cite{Boet-2003,Raymond-2012}. In practice, $\tau=1.8$ has
been used. The value $T_{E}$ is chosen as
$T_{E}=t_{0}(50+(N/5)^{3})$ where $t_{0}\sim\mathcal {O}(10^{3})$.
We run for small system a fixed large number of trials, which
increases asymptotically as $\mathcal {O}(N^{3})$ for larger
systems. This is exactly the scaling used for the Edward Anderson
problem~\cite{Boet-2001} and also working well for the next nearest
neighbor Ising problem~\cite{Raymond-2012}. In practice, the EO is
first run $X$ times (by default, $X_{0}=3$) from new random initial
conditions; if the ground state energy on run $r$ is smaller than
that found on all previous runs ($1,2,\ldots,r-1$), then $X$ is
reset as $X=X_{0}+2r$~\cite{Boet-2001,Boet-2003}. In this way, we
found that EO searches the space very efficiently from arbitrary
initial conditions, and we can have high confidence in the values
obtained.

EO not only reaches the ground state once, but resamples different
ground states many times. From these visits we can identify which
variables take the same state in a large collection of ground states
(frozen variables). To calculate the frozen set of variables we
followed in spirit the method laid out by Boettcher and
Percus~\cite{Boet-2004} for sampling a range of states with EO; our
problem is easier given that the symmetry between spin-states is
broken. To do this we allow a sequence of $O(N^{3.5})$ updates from
a random initial condition. In this extended routine, we measure the
number of frozen variables, the number of variables that take always
the same state in every ground state. When the first ground state is
visited, this number is $N$, then as subsequent ground states are
sampled, this value will decrease to its correct asymptotic value.
To implement this calculation, we simply store the marginal
magnetization $m_{i}=\left<s_{i}\right>$ averaged over visits to the
ground states, for every variable. The EO is run at a particular
value of $\tau$ and a fixed number of iterations $T_{F}$ ($\sim
10^{3}(50+(N/5)^{3.5})$), with the following additional procedures:
\begin{itemize}
  \item If the new energy of the system is smaller than the current
  ground state
  estimate, reset the marginal magnetization to the current configuration, and the size of the frozen set to $N$;
  \item If the new energy of the system is equal to the current
  ground state
  estimate, update magnetizations, and the size of the frozen set is reduced by one once observing a spin whose magnetization is no longer $\pm1$.
\end{itemize}
The above procedure provides an upper bound on the true number of
frozen variables at the identified ground state energy level. When
$\tau$ is well chosen and $T_{F}$ is large enough, it will identify
all frozen variables. As for the case of energy, we run the
algorithm $X_{0}$ times: if on a run $r$ we reduce the ground state
energy or decrease the set of frozen variables, we reset
$X=X_{0}+2r$. The set of frozen variables can be systematically
improved with each run; to do this we use the current ground state
energy estimate and current list of frozen variables as the initial
estimate for subsequent runs.

The above procedure can also be used to evaluate the entropy value.
We must now keep a record of the set of distinct ground states
visited ($\Omega_{gs}$), and only when a new ground state is visited
do we append this set. We first set the number of ground states to
$1$ and store the current configuration as a ground state in the set
$\Omega_{gs}$. If we visit all ground states we know
$s=\ln(|\Omega_{gs}|)/N$. The EO is run at a particular value of
$\tau$ and a fixed number of iterations $T_{S}$ (since the time to
discover frozen variables is precisely the time required to sample a
large fraction of the ground state space, we chose $T_{S}=T_{F}$),
with the following additional procedures:
\begin{itemize}
  \item If the new energy of the system is smaller than the current
  ground state
  estimate, reset $\Omega_{gs}$ to include only the current
  configuration, and reset the number of the ground states to one;
  \item If the new energy of the system is equal to the current
  ground state
  estimate, check whether the current configuration is in
  $\Omega_{gs}$. If not, the number of ground states increases by
  one and the current configuration is added into $\Omega_{gs}$.
\end{itemize}
In practice, we run the algorithm $X_{0}$ times and if on run $r$ we
decrease the ground state energy, or increase $|\Omega_{gs}|$, we
reset $X=X_{0}+2r$. The set $\Omega_{gs}$ is systematically improved
with each run by using the current ground state energy estimate and
$\Omega_{gs}$ as the initial estimates for subsequent runs. If
$\tau$ is well chosen and $T_{S}$ is large enough, we sample a
significant fraction of all ground states on a single run, and hence
a lower bound for the true entropy is obtained. However, for our
system we anticipate and observe extensive entropy, but storing and
comparing $O(\exp(N s))$ unique ground states is a fundamental
limitation on the size of systems we can explore. We also observe
that the time required to visit all ground states grows exponential
with system size, however $T_{S}$ is chosen large enough that for
the systems studied this is not a limitation.

It is important to note that the freezing and entropic results do
not claim to produce
 unbiased samples of the frozen set of variables, and ground states. In order to calculate the entropy of the ground state, or the set of
frozen variables, it is simply necessary to enumerate the ground
states and fair sampling is sufficient but not necessary. Simple
examples can be constructed to show that EO is a biased sampler of
ground states, thus convergence to the correct sets will be
non-uniform and possibly inefficient. However, the empirical success
of the energetic method indicates that the algorithm is not trapped
or strongly biased, and hence such a scheme can be successful on
ensembles where the energetic method works well. As is found
empirically, with reasonable runtime for the small systems
considered.


\begin{thebibliography}{48}
\expandafter\ifx\csname
natexlab\endcsname\relax\def\natexlab#1{#1}\fi
\expandafter\ifx\csname bibnamefont\endcsname\relax
  \def\bibnamefont#1{#1}\fi
\expandafter\ifx\csname bibfnamefont\endcsname\relax
  \def\bibfnamefont#1{#1}\fi
\expandafter\ifx\csname citenamefont\endcsname\relax
  \def\citenamefont#1{#1}\fi
\expandafter\ifx\csname url\endcsname\relax
  \def\url#1{\texttt{#1}}\fi
\expandafter\ifx\csname urlprefix\endcsname\relax\def\urlprefix{URL
}\fi \providecommand{\bibinfo}[2]{#2}
\providecommand{\eprint}[2][]{\url{#2}}

\bibitem[{\citenamefont{M\'ezard and Parisi}(2001)}]{cavity-2001}
\bibinfo{author}{\bibfnamefont{M.}~\bibnamefont{M\'ezard}} \bibnamefont{and}
  \bibinfo{author}{\bibfnamefont{G.}~\bibnamefont{Parisi}},
  \bibinfo{journal}{Eur. Phys. J. B} \textbf{\bibinfo{volume}{20}},
  \bibinfo{pages}{217} (\bibinfo{year}{2001}).

\bibitem[{\citenamefont{M\'ezard and Parisi}(2003)}]{cavity-2003}
\bibinfo{author}{\bibfnamefont{M.}~\bibnamefont{M\'ezard}} \bibnamefont{and}
  \bibinfo{author}{\bibfnamefont{G.}~\bibnamefont{Parisi}},
  \bibinfo{journal}{J. Stat. Phys} \textbf{\bibinfo{volume}{111}},
  \bibinfo{pages}{1} (\bibinfo{year}{2003}).

\bibitem[{\citenamefont{M\'ezard and Montanari}(2009)}]{MM-2009}
\bibinfo{author}{\bibfnamefont{M.}~\bibnamefont{M\'ezard}} \bibnamefont{and}
  \bibinfo{author}{\bibfnamefont{A.}~\bibnamefont{Montanari}},
  \emph{\bibinfo{title}{Information, Physics, and Computation}}
  (\bibinfo{publisher}{Oxford University Press}, \bibinfo{address}{Oxford},
  \bibinfo{year}{2009}).

\bibitem[{\citenamefont{Wong and Saad}(2006)}]{Wong-2006}
\bibinfo{author}{\bibfnamefont{K.~Y.~M.} \bibnamefont{Wong}} \bibnamefont{and}
  \bibinfo{author}{\bibfnamefont{D.}~\bibnamefont{Saad}},
  \bibinfo{journal}{Phys. Rev. E} \textbf{\bibinfo{volume}{74}},
  \bibinfo{pages}{010104(R)} (\bibinfo{year}{2006}).

\bibitem[{\citenamefont{Yeung and Wong}(2009{\natexlab{a}})}]{Wong-09pre}
\bibinfo{author}{\bibfnamefont{C.~H.} \bibnamefont{Yeung}} \bibnamefont{and}
  \bibinfo{author}{\bibfnamefont{K.~Y.~M.} \bibnamefont{Wong}},
  \bibinfo{journal}{Phys. Rev. E} \textbf{\bibinfo{volume}{80}},
  \bibinfo{pages}{021102} (\bibinfo{year}{2009}{\natexlab{a}}).

\bibitem[{\citenamefont{Yeung and Wong}(2009{\natexlab{b}})}]{Wong-09jstat}
\bibinfo{author}{\bibfnamefont{C.~H.} \bibnamefont{Yeung}} \bibnamefont{and}
  \bibinfo{author}{\bibfnamefont{K.~Y.~M.} \bibnamefont{Wong}},
  \bibinfo{journal}{J. Stat. Mech} \textbf{\bibinfo{volume}{P03029}}
  (\bibinfo{year}{2009}{\natexlab{b}}).

\bibitem[{\citenamefont{Yeung and Wong}(2010)}]{Wong-2010}
\bibinfo{author}{\bibfnamefont{C.~H.} \bibnamefont{Yeung}} \bibnamefont{and}
  \bibinfo{author}{\bibfnamefont{K.~Y.~M.} \bibnamefont{Wong}},
  \bibinfo{journal}{J. Stat. Mech} \textbf{\bibinfo{volume}{P04017}}
  (\bibinfo{year}{2010}).

\bibitem[{\citenamefont{Yeung and Saad}(2012)}]{CH-12}
\bibinfo{author}{\bibfnamefont{C.~H.} \bibnamefont{Yeung}} \bibnamefont{and}
  \bibinfo{author}{\bibfnamefont{D.}~\bibnamefont{Saad}},
  \bibinfo{journal}{Phys. Rev. Lett} \textbf{\bibinfo{volume}{108}},
  \bibinfo{pages}{208701} (\bibinfo{year}{2012}).

\bibitem[{\citenamefont{Rardin}(1998)}]{Rardin-98}
\bibinfo{author}{\bibfnamefont{R.~L.} \bibnamefont{Rardin}},
  \emph{\bibinfo{title}{Optimization in Operations Research}}
  (\bibinfo{publisher}{Prentice-Hall}, \bibinfo{address}{Englewood Cliffs, New
  Jersey}, \bibinfo{year}{1998}).

\bibitem[{\citenamefont{Zhou}(2005{\natexlab{a}})}]{ZH-prl05}
\bibinfo{author}{\bibfnamefont{H.}~\bibnamefont{Zhou}}, \bibinfo{journal}{Phys.
  Rev. Lett} \textbf{\bibinfo{volume}{94}}, \bibinfo{pages}{217203}
  (\bibinfo{year}{2005}{\natexlab{a}}).

\bibitem[{\citenamefont{Zhou}(2005{\natexlab{b}})}]{ZH-njp05}
\bibinfo{author}{\bibfnamefont{H.}~\bibnamefont{Zhou}}, \bibinfo{journal}{New
  J. Phys} \textbf{\bibinfo{volume}{7}}, \bibinfo{pages}{123}
  (\bibinfo{year}{2005}{\natexlab{b}}).

\bibitem[{\citenamefont{Zhou and Zhou}(2009)}]{ZH-pre09}
\bibinfo{author}{\bibfnamefont{J.}~\bibnamefont{Zhou}} \bibnamefont{and}
  \bibinfo{author}{\bibfnamefont{H.}~\bibnamefont{Zhou}},
  \bibinfo{journal}{Phys. Rev. E} \textbf{\bibinfo{volume}{79}},
  \bibinfo{pages}{020103(R)} (\bibinfo{year}{2009}).

\bibitem[{\citenamefont{Huang and Zhou}(2009)}]{Huang-pre09}
\bibinfo{author}{\bibfnamefont{H.}~\bibnamefont{Huang}} \bibnamefont{and}
  \bibinfo{author}{\bibfnamefont{H.}~\bibnamefont{Zhou}},
  \bibinfo{journal}{Phys. Rev. E} \textbf{\bibinfo{volume}{80}},
  \bibinfo{pages}{056113} (\bibinfo{year}{2009}).

\bibitem[{\citenamefont{Altarelli et~al.}(2009)\citenamefont{Altarelli,
  Braunstein, Realpe-Gomez, and Zecchina}}]{Zecchina-09}
\bibinfo{author}{\bibfnamefont{F.}~\bibnamefont{Altarelli}},
  \bibinfo{author}{\bibfnamefont{A.}~\bibnamefont{Braunstein}},
  \bibinfo{author}{\bibfnamefont{J.}~\bibnamefont{Realpe-Gomez}},
  \bibnamefont{and} \bibinfo{author}{\bibfnamefont{R.}~\bibnamefont{Zecchina}},
  \bibinfo{journal}{J. Stat. Mech} \textbf{\bibinfo{volume}{P07002}}
  (\bibinfo{year}{2009}).

\bibitem[{\citenamefont{Boettcher and Percus}(2001)}]{Boet-2001}
\bibinfo{author}{\bibfnamefont{S.}~\bibnamefont{Boettcher}} \bibnamefont{and}
  \bibinfo{author}{\bibfnamefont{A.~G.} \bibnamefont{Percus}},
  \bibinfo{journal}{Phys. Rev. Lett} \textbf{\bibinfo{volume}{86}},
  \bibinfo{pages}{5211} (\bibinfo{year}{2001}).

\bibitem[{\citenamefont{Revelle et~al.}(1977)\citenamefont{Revelle, Bigman,
  Schilling, Cohon, and Church}}]{Revell-1977}
\bibinfo{author}{\bibfnamefont{C.}~\bibnamefont{Revelle}},
  \bibinfo{author}{\bibfnamefont{D.}~\bibnamefont{Bigman}},
  \bibinfo{author}{\bibfnamefont{D.}~\bibnamefont{Schilling}},
  \bibinfo{author}{\bibfnamefont{J.}~\bibnamefont{Cohon}}, \bibnamefont{and}
  \bibinfo{author}{\bibfnamefont{R.}~\bibnamefont{Church}},
  \bibinfo{journal}{Health Services Res.} \textbf{\bibinfo{volume}{(Summer
  1977)}}, \bibinfo{pages}{129} (\bibinfo{year}{1977}).

\bibitem[{\citenamefont{Revelle}(1986)}]{Revell-1986}
\bibinfo{author}{\bibfnamefont{C.}~\bibnamefont{Revelle}},
  \bibinfo{journal}{Journal of Regional Science} \textbf{\bibinfo{volume}{26}},
  \bibinfo{pages}{343} (\bibinfo{year}{1986}).

\bibitem[{\citenamefont{Al-Karaki and Kamal}(2004)}]{Karaki-2004}
\bibinfo{author}{\bibfnamefont{J.}~\bibnamefont{Al-Karaki}} \bibnamefont{and}
  \bibinfo{author}{\bibfnamefont{A.}~\bibnamefont{Kamal}},
  \bibinfo{journal}{IEEE Wireless Communications}
  \textbf{\bibinfo{volume}{11}}, \bibinfo{pages}{6} (\bibinfo{year}{2004}).

\bibitem[{\citenamefont{Frey et~al.}(2009)\citenamefont{Frey, R\"uhrup, and
  Stojmenovi\'c}}]{Frey-2009}
\bibinfo{author}{\bibfnamefont{H.}~\bibnamefont{Frey}},
  \bibinfo{author}{\bibfnamefont{S.}~\bibnamefont{R\"uhrup}}, \bibnamefont{and}
  \bibinfo{author}{\bibfnamefont{I.}~\bibnamefont{Stojmenovi\'c}},
  \emph{\bibinfo{title}{Guide to Wireless Sensor Networks}}
  (\bibinfo{publisher}{Springer}, \bibinfo{address}{London},
  \bibinfo{year}{2009}), \bibinfo{note}{ed. by S. Misra, S.C. Misra, and I.
  Woungang}.

\bibitem[{\citenamefont{Pirkul and Jayaraman}(1998)}]{Pirkula-1998}
\bibinfo{author}{\bibfnamefont{H.}~\bibnamefont{Pirkul}} \bibnamefont{and}
  \bibinfo{author}{\bibfnamefont{V.}~\bibnamefont{Jayaraman}},
  \bibinfo{journal}{Computers and Oper. Res.} \textbf{\bibinfo{volume}{25}},
  \bibinfo{pages}{869} (\bibinfo{year}{1998}).

\bibitem[{\citenamefont{Banavar et~al.}(2000)\citenamefont{Banavar, Colaiori,
  Flammini, Maritan, and Rinaldo}}]{Banavar-2000}
\bibinfo{author}{\bibfnamefont{J.~R.} \bibnamefont{Banavar}},
  \bibinfo{author}{\bibfnamefont{F.}~\bibnamefont{Colaiori}},
  \bibinfo{author}{\bibfnamefont{A.}~\bibnamefont{Flammini}},
  \bibinfo{author}{\bibfnamefont{A.}~\bibnamefont{Maritan}}, \bibnamefont{and}
  \bibinfo{author}{\bibfnamefont{A.}~\bibnamefont{Rinaldo}},
  \bibinfo{journal}{Phys. Rev. Lett} \textbf{\bibinfo{volume}{84}},
  \bibinfo{pages}{4745} (\bibinfo{year}{2000}).

\bibitem[{\citenamefont{Shao and Zhou}(2007)}]{Shao-2007}
\bibinfo{author}{\bibfnamefont{Z.}~\bibnamefont{Shao}} \bibnamefont{and}
  \bibinfo{author}{\bibfnamefont{H.}~\bibnamefont{Zhou}},
  \bibinfo{journal}{Phys. Rev. E} \textbf{\bibinfo{volume}{75}},
  \bibinfo{pages}{066112} (\bibinfo{year}{2007}).

\bibitem[{\citenamefont{Bohn and Magnasco}(2007)}]{Bohn-2007}
\bibinfo{author}{\bibfnamefont{S.}~\bibnamefont{Bohn}} \bibnamefont{and}
  \bibinfo{author}{\bibfnamefont{M.~O.} \bibnamefont{Magnasco}},
  \bibinfo{journal}{Phys. Rev. Lett} \textbf{\bibinfo{volume}{98}},
  \bibinfo{pages}{088702} (\bibinfo{year}{2007}).

\bibitem[{\citenamefont{Wong and Saad}(2007)}]{Wong-2007}
\bibinfo{author}{\bibfnamefont{K.~Y.~M.} \bibnamefont{Wong}} \bibnamefont{and}
  \bibinfo{author}{\bibfnamefont{D.}~\bibnamefont{Saad}},
  \bibinfo{journal}{Phys. Rev. E} \textbf{\bibinfo{volume}{76}},
  \bibinfo{pages}{011115} (\bibinfo{year}{2007}).

\bibitem[{\citenamefont{Weigt and Zhou}(2006)}]{ZH-pre06}
\bibinfo{author}{\bibfnamefont{M.}~\bibnamefont{Weigt}} \bibnamefont{and}
  \bibinfo{author}{\bibfnamefont{H.}~\bibnamefont{Zhou}},
  \bibinfo{journal}{Phys. Rev. E} \textbf{\bibinfo{volume}{74}},
  \bibinfo{pages}{046110} (\bibinfo{year}{2006}).

\bibitem[{\citenamefont{Zhang et~al.}(2009)\citenamefont{Zhang, Zeng, and
  Zhou}}]{Pan-2009}
\bibinfo{author}{\bibfnamefont{P.}~\bibnamefont{Zhang}},
  \bibinfo{author}{\bibfnamefont{Y.}~\bibnamefont{Zeng}}, \bibnamefont{and}
  \bibinfo{author}{\bibfnamefont{H.}~\bibnamefont{Zhou}},
  \bibinfo{journal}{Phys. Rev. E} \textbf{\bibinfo{volume}{80}},
  \bibinfo{pages}{021122} (\bibinfo{year}{2009}).

\bibitem[{\citenamefont{Rivoire et~al.}(2004)\citenamefont{Rivoire, Biroli,
  Martin, and M\'ezard}}]{Mezard-04}
\bibinfo{author}{\bibfnamefont{O.}~\bibnamefont{Rivoire}},
  \bibinfo{author}{\bibfnamefont{G.}~\bibnamefont{Biroli}},
  \bibinfo{author}{\bibfnamefont{O.~C.} \bibnamefont{Martin}},
  \bibnamefont{and} \bibinfo{author}{\bibfnamefont{M.}~\bibnamefont{M\'ezard}},
  \bibinfo{journal}{Eur. Phys. J. B} \textbf{\bibinfo{volume}{37}},
  \bibinfo{pages}{55} (\bibinfo{year}{2004}).

\bibitem[{\citenamefont{Monasson}(1995)}]{Monasson-1995prl}
\bibinfo{author}{\bibfnamefont{R.}~\bibnamefont{Monasson}},
  \bibinfo{journal}{Phys. Rev. Lett} \textbf{\bibinfo{volume}{75}},
  \bibinfo{pages}{2847} (\bibinfo{year}{1995}).

\bibitem[{\citenamefont{Montanari et~al.}(2008)\citenamefont{Montanari,
  Ricci-Tersenghi, and Semerjian}}]{Montanari-2008jstat}
\bibinfo{author}{\bibfnamefont{A.}~\bibnamefont{Montanari}},
  \bibinfo{author}{\bibfnamefont{F.}~\bibnamefont{Ricci-Tersenghi}},
  \bibnamefont{and}
  \bibinfo{author}{\bibfnamefont{G.}~\bibnamefont{Semerjian}},
  \bibinfo{journal}{J. Stat. Mech.} \textbf{\bibinfo{volume}{P04004}}
  (\bibinfo{year}{2008}).

\bibitem[{\citenamefont{M\'ezard et~al.}(2005)\citenamefont{M\'ezard,
  Palassini, and Rivoire}}]{Mezard-2005prl}
\bibinfo{author}{\bibfnamefont{M.}~\bibnamefont{M\'ezard}},
  \bibinfo{author}{\bibfnamefont{M.}~\bibnamefont{Palassini}},
  \bibnamefont{and} \bibinfo{author}{\bibfnamefont{O.}~\bibnamefont{Rivoire}},
  \bibinfo{journal}{Phys. Rev. Lett} \textbf{\bibinfo{volume}{95}},
  \bibinfo{pages}{200202} (\bibinfo{year}{2005}).

\bibitem[{\citenamefont{Zdeborov\'a and Krzakala}(2007)}]{Lenka-2007pre}
\bibinfo{author}{\bibfnamefont{L.}~\bibnamefont{Zdeborov\'a}} \bibnamefont{and}
  \bibinfo{author}{\bibfnamefont{F.}~\bibnamefont{Krzakala}},
  \bibinfo{journal}{Phys. Rev. E} \textbf{\bibinfo{volume}{76}},
  \bibinfo{pages}{031131} (\bibinfo{year}{2007}).

\bibitem[{\citenamefont{Zhou}(2008)}]{ZH-2008}
\bibinfo{author}{\bibfnamefont{H.}~\bibnamefont{Zhou}}, \bibinfo{journal}{Phys.
  Rev. E} \textbf{\bibinfo{volume}{77}}, \bibinfo{pages}{066102}
  (\bibinfo{year}{2008}).

\bibitem[{\citenamefont{Boettcher and Percus}(2004)}]{Boet-2004}
\bibinfo{author}{\bibfnamefont{S.}~\bibnamefont{Boettcher}} \bibnamefont{and}
  \bibinfo{author}{\bibfnamefont{A.~G.} \bibnamefont{Percus}},
  \bibinfo{journal}{Phys. Rev. E} \textbf{\bibinfo{volume}{69}},
  \bibinfo{pages}{066703} (\bibinfo{year}{2004}).

\bibitem[{\citenamefont{Boettcher}(2003{\natexlab{a}})}]{Boet-2003}
\bibinfo{author}{\bibfnamefont{S.}~\bibnamefont{Boettcher}},
  \bibinfo{journal}{Phys. Rev. B} \textbf{\bibinfo{volume}{67}},
  \bibinfo{pages}{060403(R)} (\bibinfo{year}{2003}{\natexlab{a}}).

\bibitem[{\citenamefont{Raymond and Wong}(2012)}]{Raymond-2012}
\bibinfo{author}{\bibfnamefont{J.}~\bibnamefont{Raymond}} \bibnamefont{and}
  \bibinfo{author}{\bibfnamefont{K.~Y.~M.} \bibnamefont{Wong}},
  \bibinfo{journal}{J. Stat. Mech} \textbf{\bibinfo{volume}{P09007}}
  (\bibinfo{year}{2012}).

\bibitem[{\citenamefont{Boettcher}(2003{\natexlab{b}})}]{Boet-2003epjb}
\bibinfo{author}{\bibfnamefont{S.}~\bibnamefont{Boettcher}},
  \bibinfo{journal}{Eur. Phys. J. B} \textbf{\bibinfo{volume}{31}},
  \bibinfo{pages}{29} (\bibinfo{year}{2003}{\natexlab{b}}).

\bibitem[{\citenamefont{Boettcher et~al.}(2008)\citenamefont{Boettcher,
  Katzgraber, and Sherrington}}]{Boet-2008}
\bibinfo{author}{\bibfnamefont{S.}~\bibnamefont{Boettcher}},
  \bibinfo{author}{\bibfnamefont{H.~G.} \bibnamefont{Katzgraber}},
  \bibnamefont{and}
  \bibinfo{author}{\bibfnamefont{D.}~\bibnamefont{Sherrington}},
  \bibinfo{journal}{J. Phys. A} \textbf{\bibinfo{volume}{41}},
  \bibinfo{pages}{324007} (\bibinfo{year}{2008}).

\bibitem[{\citenamefont{Wei et~al.}(2012)\citenamefont{Wei, Zhang, Guo, and
  Zheng}}]{WW-2012}
\bibinfo{author}{\bibfnamefont{W.}~\bibnamefont{Wei}},
  \bibinfo{author}{\bibfnamefont{R.}~\bibnamefont{Zhang}},
  \bibinfo{author}{\bibfnamefont{B.}~\bibnamefont{Guo}}, \bibnamefont{and}
  \bibinfo{author}{\bibfnamefont{Z.}~\bibnamefont{Zheng}},
  \bibinfo{journal}{Phys. Rev. E} \textbf{\bibinfo{volume}{86}},
  \bibinfo{pages}{016112} (\bibinfo{year}{2012}).

\bibitem[{\citenamefont{Braunstein et~al.}(2005)\citenamefont{Braunstein,
  M\'ezard, and Zecchina}}]{Brau-2005}
\bibinfo{author}{\bibfnamefont{A.}~\bibnamefont{Braunstein}},
  \bibinfo{author}{\bibfnamefont{M.}~\bibnamefont{M\'ezard}}, \bibnamefont{and}
  \bibinfo{author}{\bibfnamefont{R.}~\bibnamefont{Zecchina}},
  \bibinfo{journal}{Random Struct. Algorithms} \textbf{\bibinfo{volume}{27}},
  \bibinfo{pages}{201} (\bibinfo{year}{2005}).

\bibitem[{\citenamefont{Montanari et~al.}(2007)\citenamefont{Montanari,
  Ricci-Tersenghi, and Semerjian}}]{Montanari-2007}
\bibinfo{author}{\bibfnamefont{A.}~\bibnamefont{Montanari}},
  \bibinfo{author}{\bibfnamefont{F.}~\bibnamefont{Ricci-Tersenghi}},
  \bibnamefont{and}
  \bibinfo{author}{\bibfnamefont{G.}~\bibnamefont{Semerjian}}, in
  \emph{\bibinfo{booktitle}{Proc. 45th Allerton Conf}} (\bibinfo{year}{2007}),
  pp. \bibinfo{pages}{352--359}.

\bibitem[{\citenamefont{Ricci-Tersenghi and Semerjian}(2009)}]{Tersenghi-2009}
\bibinfo{author}{\bibfnamefont{F.}~\bibnamefont{Ricci-Tersenghi}}
  \bibnamefont{and}
  \bibinfo{author}{\bibfnamefont{G.}~\bibnamefont{Semerjian}},
  \bibinfo{journal}{J. Stat. Mech} \textbf{\bibinfo{volume}{P09001}}
  (\bibinfo{year}{2009}).

\bibitem[{\citenamefont{Chavas et~al.}(2005)\citenamefont{Chavas, Furtlehner,
  M\'ezard, and Zecchina}}]{Chavas-2005}
\bibinfo{author}{\bibfnamefont{J.}~\bibnamefont{Chavas}},
  \bibinfo{author}{\bibfnamefont{C.}~\bibnamefont{Furtlehner}},
  \bibinfo{author}{\bibfnamefont{M.}~\bibnamefont{M\'ezard}}, \bibnamefont{and}
  \bibinfo{author}{\bibfnamefont{R.}~\bibnamefont{Zecchina}},
  \bibinfo{journal}{J. Stat. Mech} \textbf{\bibinfo{volume}{P11016}}
  (\bibinfo{year}{2005}).

\bibitem[{\citenamefont{Dall'Asta et~al.}(2008)\citenamefont{Dall'Asta,
  Ramezanpour, and Zecchina}}]{Zecchina-2008entropy}
\bibinfo{author}{\bibfnamefont{L.}~\bibnamefont{Dall'Asta}},
  \bibinfo{author}{\bibfnamefont{A.}~\bibnamefont{Ramezanpour}},
  \bibnamefont{and} \bibinfo{author}{\bibfnamefont{R.}~\bibnamefont{Zecchina}},
  \bibinfo{journal}{Phys. Rev. E} \textbf{\bibinfo{volume}{77}},
  \bibinfo{pages}{031118} (\bibinfo{year}{2008}).

\bibitem[{\citenamefont{Zdeborov\'a and M\'ezard}(2008)}]{Lenka-08jstat}
\bibinfo{author}{\bibfnamefont{L.}~\bibnamefont{Zdeborov\'a}} \bibnamefont{and}
  \bibinfo{author}{\bibfnamefont{M.}~\bibnamefont{M\'ezard}},
  \bibinfo{journal}{J. Stat. Mech} \textbf{\bibinfo{volume}{P12004}}
  (\bibinfo{year}{2008}).

\bibitem[{\citenamefont{Bounkong et~al.}(2006)\citenamefont{Bounkong, van
  Mourik, and Saad}}]{TABP-2006pre}
\bibinfo{author}{\bibfnamefont{S.}~\bibnamefont{Bounkong}},
  \bibinfo{author}{\bibfnamefont{J.}~\bibnamefont{van Mourik}},
  \bibnamefont{and} \bibinfo{author}{\bibfnamefont{D.}~\bibnamefont{Saad}},
  \bibinfo{journal}{Phys. Rev. E} \textbf{\bibinfo{volume}{74}},
  \bibinfo{pages}{057101} (\bibinfo{year}{2006}).

\bibitem[{\citenamefont{Gardner}(1985)}]{Gardner-1985}
\bibinfo{author}{\bibfnamefont{E.}~\bibnamefont{Gardner}},
  \bibinfo{journal}{Nucl. Phys. B} \textbf{\bibinfo{volume}{257}},
  \bibinfo{pages}{747} (\bibinfo{year}{1985}).

\bibitem[{\citenamefont{Montanari et~al.}(2004)\citenamefont{Montanari, Parisi,
  and Ricci-Tersenghi}}]{Instability-2004}
\bibinfo{author}{\bibfnamefont{A.}~\bibnamefont{Montanari}},
  \bibinfo{author}{\bibfnamefont{G.}~\bibnamefont{Parisi}}, \bibnamefont{and}
  \bibinfo{author}{\bibfnamefont{F.}~\bibnamefont{Ricci-Tersenghi}},
  \bibinfo{journal}{J. Phys. A} \textbf{\bibinfo{volume}{37}},
  \bibinfo{pages}{2073} (\bibinfo{year}{2004}).

\bibitem[{\citenamefont{Biroli and M\'ezard}(2002)}]{Biroli-02prl}
\bibinfo{author}{\bibfnamefont{G.}~\bibnamefont{Biroli}} \bibnamefont{and}
  \bibinfo{author}{\bibfnamefont{M.}~\bibnamefont{M\'ezard}},
  \bibinfo{journal}{Phys. Rev. Lett} \textbf{\bibinfo{volume}{88}},
  \bibinfo{pages}{025501} (\bibinfo{year}{2002}).

\end{thebibliography}


\end{document}